\documentclass[
]{ceurart}

\usepackage{listings}
\usepackage[colorinlistoftodos,textsize=tiny]{todonotes}
\usepackage{latexsym}
\usepackage{amssymb}
\usepackage{amsmath}
\usepackage{amsthm}
\usepackage{siunitx}
\usepackage{graphicx}
\usepackage{color}
\usepackage{import}
\usepackage{booktabs}
\usepackage{natbib}
\usepackage[ruled,vlined,linesnumbered,noresetcount]{algorithm2e}
\usepackage{pdflscape}
\usepackage{bm}
\usepackage{caption}
\usepackage{subcaption}
\usepackage{enumitem}
\usepackage{pdflscape}
\usepackage{afterpage}
\usepackage{rotating}
\usepackage[capitalise]{cleveref}
\usepackage{pctsp}
\usepackage{thmtools}
\usepackage{thm-restate}
\usepackage[title]{appendix}

\setuptodonotes{inline}

\newtheorem{theorem}{Theorem}
\newtheorem{lemma}[theorem]{Lemma}
\newtheorem{corollary}[theorem]{Corollary}

\newtheorem{definition}{Definition}

\newcommand{\BibTeX}{B\kern-.05em{\sc i\kern-.025em b}\kern-.08em\TeX}

\begin{document}

\copyrightyear{2024}
\copyrightclause{Copyright for this paper by its authors.
  Use permitted under Creative Commons License Attribution 4.0
  International (CC BY 4.0).}

\conference{ATT'24: Workshop Agents in Traffic and Transportation, October 19, 2024, Santiago de Compostela, Spain}

\title{Routing on Sparse Graphs with Non-metric Costs for the Prize-collecting Travelling Salesperson Problem}

\author[1]{Patrick O'Hara}[
    email=patrick.h.o-hara@warwick.ac.uk,
    orcid=0000-0001-9600-7554,
]\cormark[1]
\author[1]{M. S. Ramanujan}[email=r.maadapuzhi-sridharan@warwick.ac.uk,orcid=0000-0002-2116-6048]
\author[1,2,3]{Theodoros Damoulas}[email=t.damoulas@warwick.ac.uk,orcid=0000-0002-7172-4829]
\address[1]{Department of Computer Science, University of Warwick, Coventry, CV4 7AL, UK}
\address[2]{Department of Statistics, University of Warwick, Coventry, CV4 7AL, UK}
\address[3]{The Alan Turing Institute, British Library, 96 Euston Road, London, NW1 2DB, UK}

\cortext[1]{Corresponding author.}

\begin{abstract}
    In many real-world routing problems, decision makers must optimise over sparse graphs such as transportation networks with non-metric costs on the edges that do not obey the triangle inequality.
    Motivated by finding a sufficiently long running route in a city that minimises the air pollution exposure of the runner,
    we study the Prize-collecting Travelling Salesperson Problem (\pctsp) on sparse graphs with non-metric costs.
    Given an undirected graph with a cost function on the edges and a prize function on the vertices, the goal of \pctsp is to find a tour rooted at the origin that minimises the total cost such that the total prize is at least some quota.
    First, we introduce heuristics designed for sparse graphs with non-metric cost functions where previous work dealt with either a complete graph or a metric cost function. 
    Next, we develop a branch \& cut algorithm that employs a new cut we call the disjoint-paths cost-cover (DPCC) cut.
    Empirical experiments on two datasets show that our heuristics can produce a feasible solution with less cost than a state-of-the-art heuristic from the literature.
    On datasets with non-metric cost functions, DPCC is found to solve more instances to optimality than the baseline cutting algorithm we compare against.
\end{abstract}


\maketitle

\section{Introduction} \label{sec:introduction}

Variants of the Travelling Salesperson Problem (TSP) and Vehicle Routing Problem frequently appear in real-world applications \cite{hottung2020neural,nebel2009fixed,pan2023h,zhang2015efficient}.
The Prize-collecting Travelling Salesperson Problem (\pctsp) is a member of the family of TSPs with Profits \cite{Feillet2005}.
Given a graph with a prize function on the vertices and a cost function on the edges, the objective of \pctsp is to minimise the total cost of a tour that starts and ends at a given root vertex such that the total prize is at least a given threshold called the quota.
Unlike the traditional TSP, the tour is not required to visit all of the vertices.
Applications of \pctsp naturally arise in a range of industrial and transportation settings.
\citeauthor{Balas1989} \cite{Balas1989} proposes \pctsp as a model for the daily operations of a steel rolling mill.
\citeauthor{Fischetti1988} \cite{Fischetti1988} minimise the total distance travelled by a vehicle collecting a quota of a product from suppliers such that the tour starts and finishes at a factory.
\citeauthor{Awerbuch1995} \cite{Awerbuch1995} consider a salesperson visiting cities to sell a quota of brushes in order to win a trip to Hawaii.
The above applications and their respective algorithms assume either: (i) every pair of vertices are connected by an edge such that the input graph is complete; (ii) the cost function is a metric function that obeys the triangle inequality; (iii) both of the previous assumptions.

\begin{figure}[t]
    \centering
    \includegraphics[width=0.6\linewidth]{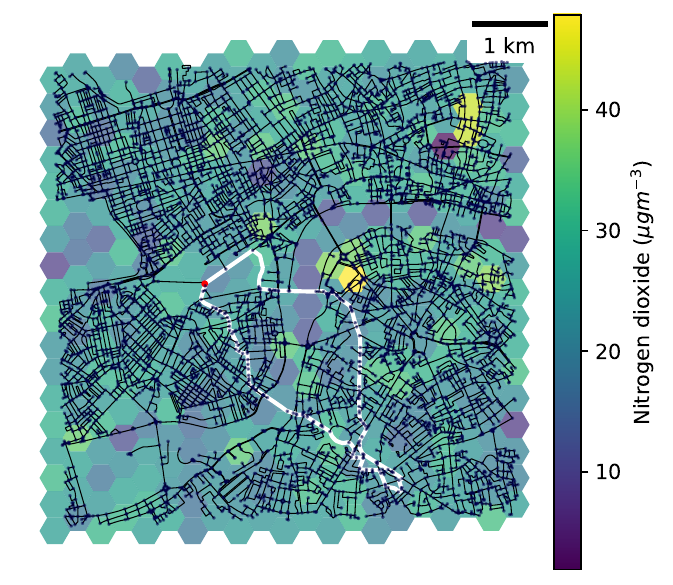}
    \caption{Example of a 10km tour that minimises the air pollution exposure of a runner in London, UK. The route starts and ends at the red vertex. Yellow cells indicate high air pollution. Blue cells indicates low air pollution.}
    \label{fig:air-quality-example}
\end{figure}

However, in many real-world applications, making assumptions that the graph is complete or the cost function is metric is not desirable.
Our motivating example is a runner planning a route through the streets of a city (see Figure~\ref{fig:air-quality-example}).
The runner requests several attributes of the route.
Firstly, the runner is conscious about exposure to air pollution on the route: air pollution has an adverse effect on the cardio-respiratory system, which can be exacerbated by increased inhalation during exercise \cite{Giles2014}.
In urban environments, air pollution is highly localised because factors such as transportation, industry and construction largely contribute to poor air quality \cite{Vardoulakis2008}. 
Fortunately, recent advances in air quality modelling~\cite{Hamelijnck2019} means live, localised air quality forecasts can be used to predict the air pollution exposure on individual streets (see \Cref{sec:results}).
Secondly, the runner requests the route starts and ends at the same location and is sufficiently long.
For example, the runner may ask for a route that starts and ends at their home and is at least 5km.
Finally, the runner asks that the route is not repetitive: whilst running up and down a single street with good air quality many times is an effective way to minimise air pollution exposure, it is also a very boring route for the runner.
We add the constraint that no section of the route is repeated, leaving other constraints on route repetition to future work.

Our motivating example can be formulated as an instance of \pctsp.
The road network is represented by an undirected graph $G$ where the edges are roads and vertices are intersections.
The road network is a sparse graph: the number of edges~$m$ is at most $\kappa n$ where $\kappa$ is a constant and $n$ is the number of vertices.
The start and end location of the route is represented by a root vertex.
The air pollution exposure on a road can be modelled by a non-metric cost function on the edges.
The length of the runner's route is analogous to prize collected on vertices by the salesperson: the prize is generated by splitting each edge $(i,j)$ in the graph into two edges $(i, k), (k, j)$ and assigning the length of edge $(i,j)$ to the prize of a newly created vertex $k$, whilst the prize of $i$ and $j$ is zero.
To avoid route repetition, we ask that the route is a simple cycle.
The objective is to minimise the air pollution exposure of a simple cycle starting and ending at the root vertex such that the length of the route is at least the given quota.

From the polyhedral results derived by \citeauthor{Balas1989} \cite{Balas1989,Balas1995,Balas2007}, three papers propose exact algorithms for \pctsp \cite{Fischetti1988,Berube2009,Pantuza2022}.
\citeauthor{Fischetti1988} \cite{Fischetti1988} derive lower bounds, however, experimental results on randomly-generated, directed, complete graphs show instances with a {\em symmetric} cost function (which is equivalent to a cost function over undirected edges) could not be solved on graphs with greater than 40 vertices due to excessive computing time.
Both \cite{Berube2009} and \cite{Pantuza2022} use cutting planes to strengthen the lower bound via valid inequalities (constraints that do not eliminate any {\em feasible} integer solutions) and conditional inequalities \cite{Fischetti1998} (constraints that are valid for every {\em optimal} solution but not for every feasible solution).
Given an upper bound on the optimal solution, cost-cover inequalities are a type of conditional inequality for \pctsp that constrain vertex variables using the cost of connecting a set of vertices.
The cost-cover inequalities of \cite{Berube2009} and \cite{Pantuza2022} both assume the cost function is metric.
Moreover, the heuristics used in \cite{Berube2009} and \cite{Pantuza2022} to obtain the upper bound assumes the graph is complete.

There are some key differences when designing polynomial-time heuristics for \pctsp on complete graphs with metric costs versus sparse graphs with non-metric costs.
First, given an instance of \pctsp on a complete graph, one can construct a feasible solution \cite{Berube2009,Pantuza2022,Pedro2013} in polynomial time; but when the graph is incomplete, we prove in \Cref{prp:feasible-heuristics} that no polynomial-time algorithm is guaranteed to recover a feasible solution to every instance of \pctsp, unless $\cP = \cN\cP$.
Second, given a tour~$\tour$ whose prize is less than the quota, heuristics \cite{Chaves2008,DellAmico1998} for complete graphs increase the prize by adding a vertex~$j$ between two adjacent vertices~$u_h, u_{h+1}$ in $\tour$ until the tour has sufficient prize; however, on sparse graphs, one cannot assume edges~$(u_h, j)$ and $(j, u_{h+1})$ always exist.
Lastly, given tour~$\tour$ whose prize is at least the quota, heuristics from the literature decrease the cost whilst maintaining feasibility by applying operations such as deleting a vertex \cite{Chaves2008}, replacing a vertex \cite{Gendreau1992,Pedro2013}, or re-sequencing the tour \cite{DellAmico1998}; similarly to above, these operators assume the existence of edges that may not exist in a sparse graph. Moreover, \cite{DellAmico1998} explicitly assumes the cost function is metric in their re-sequencing operator.
We conclude our literature review by emphasising that constructing a complete graph $G^\prime$ from our sparse input graph $G$ then naively running a heuristic from the literature on $G^\prime$ does not guarantee the solution returned by the heuristic will be feasible for $G$.
Furthermore, such a construction increases the number of edges to $\cO(n^2)$, slowing down the running time.
We give two constructions in Appendix~\ref{sec:constructions} of the supplementary material.

\paragraph{Contributions} With the motivation of finding routes minimising air pollution exposure in mind, the focus of this paper is to develop algorithms to solve \pctsp on sparse, undirected graphs with non-metric cost functions.
Our contributions are:
\begin{enumerate}
    \item Proposing a {\em three-stage heuristic} called \sblpec\xspace (SBL-PEC) which (i) generates an initial low-cost tour; (ii) increases the prize of the tour by adding paths with the smallest cost-to-prize ratio; and (iii) decreases the cost of the tour whilst maintaining feasibility by swapping a sub-path of the tour for an alternative path with less cost. Stages (ii) and (iii) contain \cite{DellAmico1998} as a special case.
    \item Deriving a new {\em disjoint-paths cost-cover inequality} and proving it is stronger than a shortest-path cost-cover inequality (\Cref{prp:cost-cover}). Our disjoint-paths cost-cover inequality is integrated into our branch \& cut algorithm for solving instances to optimality.
    \item Empirically evaluating our method on real-world instances from the London air quality dataset \cite{Hamelijnck2019} and on synthetic instances from TSPLIB \cite{Reinelt1991}. We compare our heuristics against \cite{DellAmico1998} and compare our branch \& cut algorithm against an adaptation of \cite{Berube2009}.
\end{enumerate}

\section{Problem definition} \label{sec:problem-definition}

An instance of \pctsp is given by the tuple $\cI = (G, c, p, Q, \rootvertex)$.
The graph $G$ is undirected with vertices $V(G)$ and edges $E(G)$.
The graph is assumed to be connected and simple with no self-loops.
The number of vertices and the number of edges are denoted $n=|V(G)|$ and $m=|E(G)|$ respectively.
The set of neighbours of vertex $i$ is denoted $N(i) = \{ j \in V(G) : (i, j) \in E(G) \}$.
The cost function $c:E(G) \to \mathbb{N}_0$ defined on the edges is non-negative where $\mathbb{N}_0 = \mathbb{N} \cup \{ 0 \}$.
The prize function $p:V(G) \to \mathbb{N}_0$ defined on the vertices is also non-negative.
For convenience, we denote the total prize of a set of vertices $S \subseteq V(G)$ as $p(S) = \sum_{i \in S}~{p(i)}$.
Similarly, the total cost of a set of edges $F \subseteq E(G)$ is $c(F) = \sum_{(i,j) \in F}~{c(i,j)}$.
We are given a non-negative quota $Q \in \mathbb{N}_0$.
The root vertex is $\rootvertex \in V(G)$.

\begin{definition}\label{def:tour}
    A tour $\tour = (u_1, \ldots, u_k, u_1)$ of an undirected graph $G$ is a sequence of adjacent vertices such that each vertex in the tour is visited exactly once.
    A vertex $u$ is visited exactly once if the tour enters $u$ exactly once and leaves $u$ exactly once.
    The set of vertices in the tour is defined by $V(\tour) = \{ u_1, \ldots, u_k \}$ and the set of edges of the tour is defined by $E(\tour) = \{ (u_1, u_2), \ldots, (u_{k-1}, u_k), (u_k, u_1) \}$.
    We say a tour is prize-feasible if $p(V(\tour)) \geq Q$.
\end{definition}

A {\em feasible solution} to a given instance~$\cI$ of \pctsp is a tour~$\tour$ starting and ending at the given root vertex~$v_1$ such that $\tour$ is prize-feasible.
The objective of \pctsp is to minimise the total cost~$c(E(\tour))$ of the edges in a tour~$\tour$ such that the tour is a feasible solution.
A tour~$\tour^\star$ is an {\em optimal solution} to $\cI$ if for all feasible tours~$\tour$ we have $c(E(\tour^\star)) \leq c(E(\tour))$.

\begin{definition}\label{def:sparse}
    A graph is sparse if the number of edges~$m$ is at most~$\kappa n$, where $\kappa$ is a positive constant.
\end{definition}
\begin{definition}\label{def:metric}
    Let $\cP_{uv}$ be the least-cost path from vertex~$u$ to vertex~$v$ in $G$ and let $E(\cP_{uv})$ be the set of path edges.
    An edge~$(u,v)$ is metric if $\forall w \in V(G)$: (i) $c(u,v) = 0 \Leftrightarrow u = v$, (ii) $c(u,v) = c(v,u)$, (iii) $c(u,v) \leq c(E(\cP_{uw})) + c(E(\cP_{wv}))$.
    A cost function is metric if edge~$(u,v)$ is metric for $\forall (u,v) \in E(G)$, and the cost function is non-metric otherwise.
\end{definition}

\begin{restatable}{lemma}{lemmetricedges}
\label{lem:metric-edges}
    Let $G$ be a connected, undirected graph and $c: E(G) \to \mathbb{N}_0$ be a cost function on the edges.
    The number of metric edges in $E(G)$ is at least $n-1$.
\end{restatable}

\begin{definition} \label{def:metric-surplus}
    Given graph~$G$ and cost function $c: E(G) \to \mathbb{N}_0$, the metric surplus~$\zeta(G, c)$ is:
    \begin{align*}
        \zeta(G, c) = \frac{ \left( \sum_{(u, v) \in E(G)}~{M(u, v)} \right) - (n - 1)}{m - (n - 1)}
    \end{align*}
    where $M(u, v) = 1$ if edge $(u, v)$ is metric, or 0 otherwise.
\end{definition}

In this paper, we assume the input graph is sparse and we do not assume the triangle inequality (iii) in \Cref{def:metric} always holds for the cost function.
The metric surplus from \Cref{def:metric-surplus} gives us a measure of ``how metric'' a cost function is.
Notice that if all edges in $G$ are metric, then $\zeta(G, c) = 1$.
But if the number of metric edges is equal to the lower bound $n-1$ from \Cref{lem:metric-edges}, then $\zeta(G, c) = 0$.
Finally, given an instance $\cI$, we apply pre-processing to reduce the size of a sparse input graph $G$ by removing vertices that are not in the same biconnected component as the root vertex $v_1$ (see Appendix~\ref{sec:preprocessing}).

\section{Heuristics} \label{sec:heuristics}

Heuristics for \pctsp provide an upper bound on the optimal solution.
A heuristic that runs efficiently in worst-case polynomial-time complexity is desirable.
However, on a general graph which is not guaranteed to be complete, the design of polynomial-time heuristics that always return a prize-feasible tour (if one exists) is not possible:

\begin{restatable}{proposition}{feasibleheuristics}
Assume ${\cal P} \neq \NP$. Let $G$ be any undirected graph. No polynomial-time algorithm $A$ is guaranteed to find a feasible solution (if one exists) for every instance of \pctsp.
\label{prp:feasible-heuristics}
\end{restatable}
\begin{proof}
    Reduction from Hamiltonian Cycle: see Appendix~\ref{sec:heuristics-proof}.
\end{proof}

Whilst the polynomial-time heuristics we design in this section are not guaranteed to always find a feasible solution on {\em all} general input graphs due to \Cref{prp:feasible-heuristics}, our aim is to design heuristics that find a low-cost feasible solution on {\em most} sparse, non-metric instances in practice.
This section presents a combined heuristic (\Cref{sec:path-extension-collapse}) comprising of three distinct components: generating a starting solution (\Cref{sec:suurballes}); increasing the prize of the tour (\Cref{sec:path-extension}); and reducing the cost of the tour (\Cref{sec:path-collapse}).

\subsection{Suurballe's Heuristic} \label{sec:suurballes}

To find an initial low-cost tour, we propose \sbl\xspace (SBL).
The central idea is to find $n-1$ tours $\tour_1, \ldots, \tour_{n-1}$ from the least-cost pair of vertex-disjoint paths from the root vertex $v_1$ to every other vertex $t \in V(G) \backslash \{ v_1 \}$.
Two simple paths $\cD_1 = (u_1, \ldots, u_k)$, $\cD_2 = (w_1, \ldots, w_l)$ from $v_1 = u_1 = w_1$ to $t = u_k = w_l$ are {\em vertex disjoint} if $V(\cD_1) \cap V(\cD_2) = \{ v_1, t \}$.
In an undirected graph, $\cD_1$ and $\cD_2$ form a tour $\cT = (u_1, \ldots, u_{k-1}, u_k, w_{l-1}, \ldots, w_1)$ by removing $w_l$ from $\cD_2$, reversing the order of $\cD_2$, then appending the modified $\cD_2$ to the end of $\cD_1$.


\begin{figure}[t]
    \centering 
    \begin{subfigure}{0.47\textwidth}
        \centering
        \includegraphics[width=\linewidth]{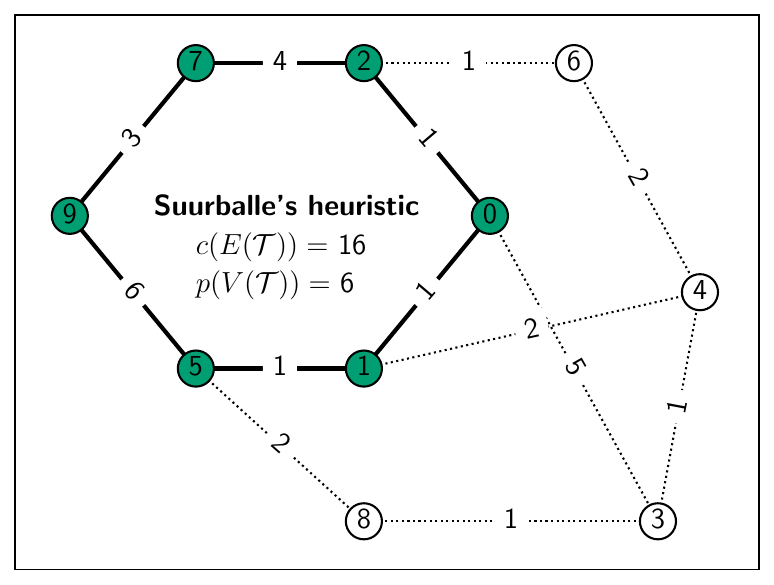}
        \caption{\sbl\xspace generates the initial tour (0,1,5,9,7,2,0) from a pair of vertex-disjoint paths (0,1,5,9) and (0,2,7,9). Vertices in green are in the tour. The total cost of $\cT$ is 16 and the total prize is 6. We set $\tour^\star \leftarrow \tour$.}
        \label{fig:sh}
    \end{subfigure}\hfil 
    \begin{subfigure}{0.47\textwidth}
        \centering
        \includegraphics[width=\linewidth]{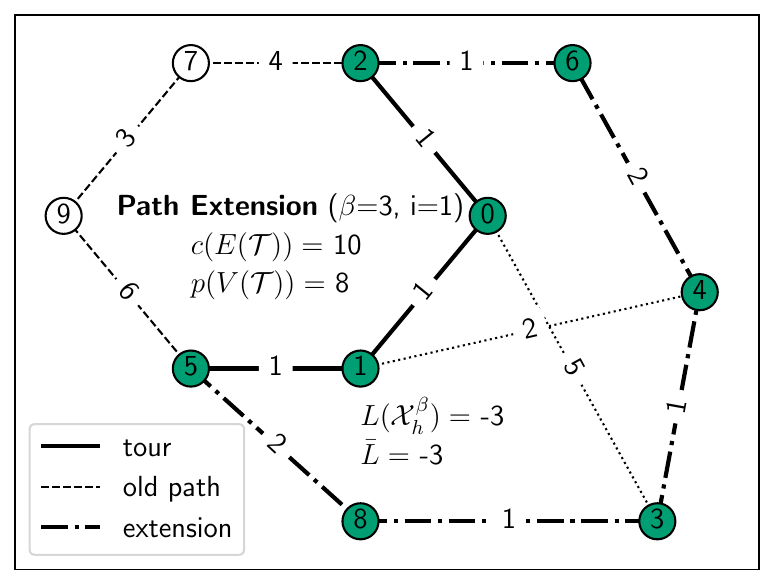}        
        \caption{Path extension of \cref{fig:sh} with $\beta = 3$ finds a tour with cost 10 and prize 8. 
        The only feasible extension on iteration $i=1$ is (2,6,4,3,8,5). 
        For~$i=2$, path (2,7,9,5) is not considered because its prize is not greater than the internal path (2,0,1,5).}
        \label{fig:pe3}
    \end{subfigure}
    \caption{Demonstration of the SBL and \pe\xspace stages of PEC-SBL heuristic (Algorithm~1). Vertex $0$ is the root. The prize of every vertex is 1. Edges are labelled with their cost. The quota is $Q=6$.}
    \label{fig:step-by-step}
\end{figure}
SBL returns the least-cost prize-feasible tour from $\tour_1, \ldots, \tour_{n-1}$ if one exists; else if no tours are prize-feasible, it returns the tour that maximises the prize.
Suurballe's algorithm\footnote{Suurballe's algorithm uses directed, asymmetric graphs. Transforming undirected to directed, asymmetric takes $\cO(n + m)$.} \cite{Suurballe1984} for finding the least-cost pair of vertex-disjoint paths from $v_1$ to every other vertex $t$ takes $\cO(m \log n)$.
Checking a single tour is prize-feasible takes $\cO(n)$ and the heuristic checks $n-1$ tours for prize-feasibility.
The overall running time of our algorithm is $\cO(m \log n + n^2)$.
On a sparse graph, the worst-case is dominated by $\cO(n^2)$.
On a dense graph, the worst-case is dominated by $\cO(m \log n)$.
The weakness of SBL is it fails to find a prize-feasible tour if there does not exist a least-cost pair of vertex-disjoint paths $\cD_1, \cD_2$ from $\rootvertex$ to some $t \in V(G)$ such that $p(V(\cD_1)) + p(V(\cD_2)) - p(t) \geq Q$.

\subsection{Path Extension} \label{sec:path-extension}

Given an initial tour~$\tour = (u_1, \ldots, u_k, u_1)$ with $u_1 = \rootvertex$, we propose a \pe\xspace algorithm to increase the total prize of the tour.
Increasing the prize has two goals: first, if $\tour$ is not prize-feasible, then increase the prize to obtain feasibility; second, if $\tour$ is prize-feasible, then explore the solution space in different areas of the graph by increasing the prize.
Intuitively, in each iteration $i$, the algorithm searches for possible {\em extension paths} between two vertices of the tour, then chooses the extension path with the smallest cost-to-prize ratio.
A new tour with larger prize is constructed from the extension path and the existing tour.
The algorithm repeatedly increases the prize of the tour with extension paths until a termination criteria is met.

More specifically in each iteration $i$, for every $h \in \{ 1, \ldots, k - \beta \}$, \pe\xspace searches for an extension path $\cX_h^\beta$ from $u_h$ to $u_{h + \beta}$ where $\beta \in \mathbb{N}$ is a step size parameter and $k = |V(\tour)|$.
The extension path $\cX_h^\beta$ is not allowed to use vertices in $\tour$ except for $u_h$ and $u_{h + \beta}$.
To find $\cX_h^\beta$, we use a Breadth First Search (BFS) from $u_h$ such that $|V(\cX_h^\beta)| \geq 3$.
Let us define the {\em internal path}~$\cP_h^\beta = (u_h, u_{h+1}, \ldots, u_{h+\beta})$ to be the sub-path of $\tour$ that would be replaced by the extension path~$\cX_h^\beta$.
A new tour can be created by replacing the internal path with the extension path.
To compare extension paths, we propose a ratio of the cost to prize called the {\em unitary loss}:
\begin{align}
    L(\cX_h^\beta) &= \frac{c(E(\cX_h^\beta)) - c(E(\cP_h^\beta))}{p(V(\cX_h^\beta)) - p(V(\cP_h^\beta))} & \text{where}~p(V(\cX_h^\beta)) > p(V(\cP_h^\beta)) \label{eq:unitary-loss}
\end{align}

In each iteration $i$, we construct a set~$S_i^\beta = \{ \cX_h^\beta | p(V(\cX_h^\beta)) > p(V(\cP_h^\beta)),  h \in \{ 1, \ldots, k - \beta \} \}$ of possible extension paths that have more prize than the internal path.
Note we exclude from $S_i^\beta$ extension paths $\cX_h^\beta$ for which $h > k - \beta$ because removing the internal path $\cP_h^\beta$ would also remove the root vertex from the tour.
We choose the extension path $\cX_h^\beta \in S_i^\beta$ that minimises $L(\cX_h^\beta)$ and construct a new tour~$\tour^\prime = (u_1, \ldots, u_{h-1}, \cX_h^\beta, u_{h + \beta + 1}, \ldots, u_k, u_1)$.
We repeat steps for at most $n$ iterations or until termination criteria (a) or (b) is met: 
\begin{enumerate}[label=(\alph*)]
    \item If the prize of the initial tour is less than $Q$, \pe\xspace terminates when the tour becomes prize-feasible, or terminates when $S_i^\beta$ is empty (the heuristic failed to find a prize-feasible tour).
    \item If the prize of the initial tour is at least $Q$, \pe\xspace terminates when there does not exist a path $\cX_h^\beta \in S_i^\beta$ such that $L(\cX_h^\beta) < \overline{L}$ where: \begin{align}\overline{L} = \frac{1}{|S_{i=1}^\beta|}\sum_{\cX_h^\beta \in S_{i=1}^\beta}~{L(\cX_h^\beta)} \end{align}
    $\overline{L}$ is calculated on the first iteration $i=1$, but not re-calculated on subsequent iterations.
\end{enumerate}
Each iteration $i$ takes $\cO(n(n + m))$ time, and there are at most $n$ iterations, giving \pe\xspace worst-case time complexity $\cO(n^2(n + m))$.
We found through experimentation that if $\beta > 10$, the tour is rarely extended, thus we limit $\beta$ to be at most $\beta_{\max} = 10$.
We limit the maximum number of iterations to be $n$ because the worst-case running time would otherwise be upper bounded by the number of times one can increase the prize of the tour by replacing a sub-path with an extension path.
Depending on the prize function and topology of the graph, this upper bound could be much larger than $n$.
Since we seek a fast heuristic (and since \pe\xspace will be called $\beta_{\max}$ times, see \Cref{sec:path-extension-collapse}), we limit the maximum number of iterations to be $n$.

The \pe\xspace algorithm addresses limitations of previous work in the following ways.
First, on sparse graphs, \pe\xspace can increase the prize of the tour by adding paths between non-adjacent vertices in the tour: it is not limited by adding a single vertex between adjacent vertices of the tour.
For example, in \Cref{fig:pe3}, \pe\xspace recovers the optimal solution by finding an extension path with five edges that replaces an internal path with three edges.
Second, the unitary loss function (\ref{eq:unitary-loss}) does not assume the cost function is metric.
Indeed if an edge $(u_h, u_{h+1}) \in E(\tour)$ is non-metric such that the cost of $(u_h, u_{h+1})$ is greater than the cost of the extension path $\cX_h^1$, then $L(\cX_h^1) < 0$ so \pe\xspace is likely to choose $\cX_h^1$ as the extension path.
In this case, extending the tour with $\cX_h^1$ decreases the cost and increases the prize of the tour, which is a desirable greedy move for the algorithm to make.



\subsection{Path Collapse} \label{sec:path-collapse}

To reduce the cost of a given prize-feasible tour $\tour = (u_1, \ldots, u_k, u_1)$, we propose the \pc\xspace heuristic.
The intuition is to swap a sub-path of the tour for an alternative path with less cost whilst maintaining a prize-feasible tour containing the root vertex.
We iterate over $k$ sub-paths of the tour (where $k = |V(\tour)|$) and choose the alternative path that produces the least-cost prize-feasible tour.
More specifically, \pc\xspace iterates over each $i \in \{ 1, \ldots, k \}$:
\begin{enumerate}
    \item Start from vertex $u_i$, then take a sub-path $P = (u_i, \ldots, u_j)$ of $\tour$ where $v_1 \in V(P)$ such that $p(V(P)) < Q$ and $p(V(P)) + p(u_{j+1}) \geq Q$.
    \item For all neighbours~$s_l \in N(u_i)$ of $u_i$, search for the least-cost path~$\cS_l = (s_1, \ldots, s_l)$ from $s_1=u_j$ to $s_l$ using vertices of $G$ that are not in $P$.
    \item From $s_l \in N(u_i)$, choose the path $\cS_l^\star$ that minimises $c(E(\cS_l)) + c(s_l, u_j)$ such that $p(V(P)) + p(V(\cS_l)) - p(u_j) \geq Q$. If $c(E(P)) + c(E(\cS_l^\star)) + c(s_l, u_j) < c(E(\tour))$, then form a new collapsed tour $\tour_i^\prime = (u_i, \ldots, u_j, s_2, \ldots, s_l, u_i)$ by appending $\cS_l^\star$ and $(s_l, u_i)$ to $P$.
\end{enumerate}
From the $k$ collapsed tours $\tour_i^\prime$ returned by step 3 in each iteration $i \in \{ 1, \ldots, k \}$, \pc\xspace returns the least-cost tour.
Using Dijkstra's algorithm \cite{Dijkstra1959} for shortest paths, the running-time of \pc\xspace is $\cO(k \cdot m \log n)$.
Note that our \pc\xspace algorithm contains \cite{DellAmico1998} as a special case when the shortest path from $u_j$ to $u_i$ has exactly two edges.
\pc\xspace is a generalization that allows shortest paths with more than two edges.
This generalization is critical in a sparse graph when a two-edge path from $u_j$ to $u_i$ may not exist.
If edge $(u_j, s_l)$ exists and is non-metric, then the generalization is important because there likely exists a least-cost path $\cS_l$ from $u_j$ to $s_l \in N(u_j)$ that does not visit edge $(u_j, s_l)$ such that the resulting tour $\tour_i^\prime$ has less cost and more prize (since $p(u_j) + p(s_l) < p(\cS_l)$).

\subsection{Suurballe's Path Extension \& Collapse (SBL-PEC)} \label{sec:path-extension-collapse}
\begin{algorithm}[h]
    \caption{Path extension \& collapse initialised with Suurballe's heuristic (SBL-PEC).}
    \KwIn{Instance $\cI = (G, c, p, Q, \rootvertex)$; parameter $\beta_{\max} \in \mathbb{N}$. \hspace{1em} \textbf{Output:} Tour $\tour^\star$.}
    Find a tour $\cT$ using SBL \;
    \lIf{$p(V(\tour)) < Q$}{\textbf{foreach} $\beta \in \{ 1, \ldots, \beta_{\max}\}$ \textbf{do} \pe\xspace with termination \textbf{(a)}}
    $\cT^\star \leftarrow $ \pc\xspace $\cT$\;
    \For{$\beta \in \{ 1, \ldots, \beta_{\max} \}$} {
        $\tour \leftarrow $ \pe\xspace on $\cT^\star$ with step size $\beta$ with termination criteria \textbf{(b)}\;
        $\tour \leftarrow $ \pc\xspace $\tour$ \;
        \lIf{$c(E(\tour)) < c(E(\tour^\star))$}{$\tour^\star \leftarrow \tour$}
    }
\end{algorithm} 

To summarise SBL-PEC (Algorithm~1), we generate an initial tour $\cT$ with SBL; if $\cT$ is not prize-feasible, we repeat \pe\xspace for each $\beta$ until $p(V(\cT)) \geq Q$; then for each $\beta$, we alternate \pe\xspace with \pc.
SBL-PEC has a worst-case time complexity equal to the number of times we call \pe, which is $\cO(\beta_{\max} n^2(n+m))$.

\section{Branch \& Cut Algorithm} \label{sec:branch-cut}

In this section, we develop a branch \& cut algorithm to find the optimal solution of sparse, non-metric instances of \pctsp.
Our primary contribution is a new conditional cost-cover inequality based on vertex-disjoint paths (\cref{sec:cost-cover}).
For a computational study on the effectiveness of other valid inequalities for \pctsp such as cover and comb inequalities, we refer the reader to \cite{Berube2009}.
Our branch \& cut algorithm is implemented using SCIP v8.0.3 \cite{Achterberg2008,Achterberg2009,Bestuzheva2021} and linear programs are solved using CPLEX v22.1.1 \cite{cplex2009}.

\subsection{Integer Programming Formulation}

We formulate \pctsp as an integer linear program (ILP) as follows \cite{Berube2009}.
Let variable $y_i = 1$ if vertex~$i$ is included in the tour, or zero otherwise.
Similarly, let $x_{ij} = 1$ if edge~$(i,j)$ is included in the tour, or zero otherwise.
We can summarise the variables as binary vectors $\bm{x} \in \{ 0, 1 \}^m$ and $\bm{y} \in \{ 0, 1 \}^n$.
Similarly the cost and prize can be represented by vectors $\bm{c} \in \mathbb{N}_0^m$ and $\bm{p} \in \mathbb{N}_0^n$ respectively.
Our ILP is
\begin{align}
    & \min & & \bm{c}^T\bm{x}& \label{eq:objective-function} \\
    & \text{s.t.} & & \bm{p}^T\bm{y} \geq Q&& \label{con:prize} \\
    & & & y_1 = 1 && \label{con:root-vertex} \\
    & & & \sum_{j \in N(i)}~{x_{ij}} = 2 y_i && \forall i \in V(G) \label{con:degree} \\
    & & & \sum_{(j,k) \in E(S)}~{x_{jk}} \leq \sum_{j \in S}~{y_j} - y_i && \forall S \subset V(G),\hspace{0.5em} \rootvertex \in V(G) \backslash S, \hspace{0.5em}  v_i \in S \label{con:sub-tour-elimination} \\
    & & & \bm{x} \in \{ 0, 1 \}^m,\hspace{0.5em} \bm{y} \in \{ 0, 1 \}^n. && \label{con:the-rest}
\end{align}
Constraint (\ref{con:prize}) enforces the total prize of the tour to be at least the quota.
Constraint (\ref{con:root-vertex}) requires the root vertex to be in every feasible solution.
Constraint (\ref{con:degree}) ensures every vertex in the graph is visited at most once by the tour.
Constraints (\ref{con:sub-tour-elimination}) are called the sub-tour elimination constraints (SECs) which ensure the tour is connected.
Constraints \eqref{con:the-rest} requires the edge and vertex variables to be binary.
The set of feasible solutions is given by $\cF = \{ (\bm{x}, \bm{y}) \in \{ 0, 1 \}^m \times \{ 0, 1 \}^n | (\ref{con:prize})-(\ref{con:the-rest}) \}$.
A solution $(\bm{x}^\star, \bm{y}^\star) \in \cF$ is optimal if for all $(\bm{x}, \bm{y}) \in \cF$ we have $\bm{c}^T \bm{x}^\star \leq \bm{c}^T \bm{x}$.
A relaxation is obtained by removing some of the constraints from the integer program and can be used to obtain a lower bound (LB) on the cost of the optimal solution.
Given an upper bound (UB) on the optimal solution, we define the gap between the bounds as GAP = (UB - LB) / LB.
A solution is optimal when the GAP is zero.

The linear program at the root node of the branch \& bound tree is defined by the objective function (\ref{eq:objective-function}) with constraints (\ref{con:prize})-(\ref{con:degree}).
The integrality constraints on $\bm{x}$ and $\bm{y}$ are relaxed to $\forall (i,j) \in E(G): 0 \leq x_{ij} \leq 1$ and $\forall i \in V(G): 0 \leq y_i \leq 1$.
Sub-tour elimination constraints are only added to a linear program when a separation algorithm finds a violated inequality (Appendix~\ref{sec:sec}). 
The initial upper bound $C_U$ is set by running the SBL-PEC heuristic from \Cref{sec:heuristics}.
If the value of the objective function of a solved linear program is greater than the current upper bound, the node belonging to the linear program is pruned from the branch \& bound tree.

\subsection{Cost-cover Inequalities} \label{sec:cost-cover}

Given an upper bound $C_U$ on the optimal solution of an instance of \pctsp and a subset $S$ of vertices, a cost-cover inequality constrains the number of vertices from $S$ that can be in the optimal solution.
Let $l: S \to \mathbb{N}_0$ be a function that returns a lower bound on the cost of connecting vertices in $S$ with any edges from $E(G)$.
Note that $C_U$ acts as the capacity in a knapsack constraint: $\sum_{(i,j) \in E(G)}~{x_{ij}c(i,j)} \leq C_U$.
Under the {\em condition} $l(S) > C_U$, $S$ is a cover of the knapsack constraint, and no optimal solution contains every vertex from $S$.
The following conditional inequality is then valid for the optimal solution (but not for every feasible solution):
\begin{align}
& \sum_{i \in S}~{y_i} \leq |S| - 1 & \forall S \subset V(G),\hspace{1em} l(S) > C_U \label{eq:cost-cover}
\end{align}
Consider the case when $S$ contains two vertices~$i$ and $j$.
On a non-metric and sparse graph, if twice the cost of the shortest path $\cP_{ij}$ between $i$ and $j$ is greater than~$C_U$, then at most one of the endpoints~$i,j$ can be included in the optimal solution.
Formally, the {\em shortest-path cost-cover} (SPCC) inequality is
\begin{align}
    & y_i + y_j \leq 1 & \forall i,j \in V(G),\hspace{1em} 2 c(E(\cP_{ij})) > C_U. \label{eq:shortest-path-cost-cover}
\end{align}
We propose using the least-cost pair of vertex-disjoint paths between two vertices $i,j$ instead of two times the cost of the shortest path.
We first prove the cost of the disjoint paths are a lower bound on the shortest tour containing $i$ and $j$:
\begin{restatable}{lemma}{disjoint}
\label{lem:disjoint}
In an undirected graph, the minimum cost of a tour $\cT$ containing $i$ and $j$ is equal to the least-cost pair of vertex-disjoint paths $\cD_1, \cD_2$ between $i$ and $j$.
\end{restatable}
\begin{proof}
    See Appendix~\ref{sec:cost-cover-proofs} of the supplementary material.
\end{proof}
By \Cref{lem:disjoint}, $l(\{i, j \}) = c(E(\cD_1))+ c(E(\cD_2))$ defines a lower bound on a tour containing $i$ and $j$.
We define the {\em disjoint-paths cost-cover} (DPCC) inequality as:
\begin{align}
    & y_i + y_j \leq 1 & \forall i,j \in V(G),\hspace{1em} c(E(\cD_1))+ c(E(\cD_2)) > C_U. \label{eq:disjoint-path-cost-cover}
\end{align}
\begin{restatable}{proposition}{costcover}
\label{prp:cost-cover}
    Let $C_u$ be an upper bound on the optimal solution to an instance $\cI$.
    Let~$\cF^{(\ref{eq:shortest-path-cost-cover})}$ and $\cF^{(\ref{eq:disjoint-path-cost-cover})}$ be the set of feasible solutions given the SPCC~\eqref{eq:shortest-path-cost-cover} and DPCC~\eqref{eq:disjoint-path-cost-cover} inequalities respectively, that is,
    \begin{align*}
        \cF^{(\ref{eq:shortest-path-cost-cover})} := \{(\bm{x}, \bm{y}) \in \cF \mid (\ref{eq:shortest-path-cost-cover}) \},
        \hspace{4em}  \cF^{(\ref{eq:disjoint-path-cost-cover})} := \{(\bm{x}, \bm{y}) \in \cF \mid (\ref{eq:disjoint-path-cost-cover}) \}.
    \end{align*}
    Then we have $\cF^{(\ref{eq:disjoint-path-cost-cover})} \subseteq \cF^{(\ref{eq:shortest-path-cost-cover})}$.
\end{restatable}
\begin{proof}
    To show $\cF^{(\ref{eq:disjoint-path-cost-cover})} \subseteq \cF^{(\ref{eq:shortest-path-cost-cover})}$, we prove that if $(\bm{x}, \bm{y}) \in \cF^{(\ref{eq:disjoint-path-cost-cover})}$ then $(\bm{x}, \bm{y}) \in \cF^{(\ref{eq:shortest-path-cost-cover})}$.
    Suppose otherwise.
    Then $\exists (\bm{x}, \bm{y}) \in \fdp$ such that $(\bm{x}, \bm{y}) \notin \fsp$.
    This only happens if $\exists i, j \in V(G)$ such that $y_i + y_j \leq 1$ is a constraint in $\fsp$ but is not a constraint in $\fdp$.
    This means $2 c(E(\cP_{ij})) > C_U$ and $c(E(\cD_1)) + c(E(\cD_2)) \leq C_u$.
    Without loss of generality, let $c(E(\cD_1)) \leq c(E(\cD_2))$.
    Then $c(E(\cD_1)) \leq \frac{1}{2} C_u$ which implies $\cD_1$ is a shorter path from $i$ to $j$ than $\cP_{ij}$,
    a contradiction to (\ref{eq:shortest-path-cost-cover}) that $\cP_{ij}$ is the shortest path.
\end{proof}
Our branch \& cut algorithm checks for violated cost-cover inequalities between the root vertex $v_1$ to every other $v_i \in V(G)$.
Before running the branch \& cut algorithm, we find the least-cost pair of vertex-disjoint paths from $v_1$ to every $v_i \in V(G)$ and store the costs in an array $A$ with $n$ elements.
This pre-processing step takes $\cO(m \log n)$ for disjoint paths \cite{Suurballe1984}.
Note that we decided against finding the least-cost pair of vertex-disjoint paths between every pair of vertices because it would take $\cO(n m \log n)$ time complexity and $\cO(n^2)$ space complexity.
During the branch \& cut algorithm, whenever a new upper bound $C_U$ is discovered, if $A[i] > C_U$ then we set $y_i = 0$ due to (\ref{eq:disjoint-path-cost-cover}).
The time complexity of the separation algorithm (discarding the time for pre-processing) is $\cO(n)$ since we must check every element of the array $A$.
The cost-cover separation algorithms for shortest-path cost-cover inequalities (\ref{eq:shortest-path-cost-cover}) and disjoint-paths cost-cover inequalities (\ref{eq:disjoint-path-cost-cover}) are the same, except we store the shortest path from $v_1$ to $v_i \in V(G)$ in array $A$ using Dijkstra's algorithm \cite{Dijkstra1959}.

\section{Computational experiments} \label{sec:results}

We evaluate our heuristics and cost-cover inequalities across multiple real-world and synthetic instances.
For each instance and each algorithm, we measure the computational time in seconds (TIME) and the GAP between the upper and lower bounds.
TIME is limited to four hours for each instance.
For a given algorithm evaluated on a group of instances, we denote the mean GAP as \gap, the mean TIME as \avgtime, the number of feasible solutions as FEAS, and the number of optimal solutions as OPT.
Our algorithms are implemented in C++ and Python\footnote{Code available at https://github.com/PatrickOHara/pctsp. Datasets available upon request.}.
The machine we use for experiments is a $2 \times 10$-core Xeon E5-2660 v3 at 2.6 GHz with 64GB RAM running Linux.
Each run of an algorithm is allocated one core and 12GB of RAM using the slurm scheduler \cite{Yoo2003}.


\paragraph{London Air Quality (LAQ)}
We generate instances of the LAQ dataset by mapping air pollution forecasts from a machine learning model onto the London road network.
The air quality model of London \cite{Hamelijnck2019} is a non-stationary mixture of Gaussian Processes that predicts air pollution (nitrogen dioxide) from data such as sensors, road traffic and weather.
The model output consists of forecasts for each cell of a hexagonal grid which overlays the road network (see \Cref{fig:air-quality-example}).
Every road has a length (in meters).
The cost of running along a road is the mean pollution of the grid cells intersecting the road multiplied by the length of the road.
The goal is to minimise air pollution exposure of a tour such that the total length of the tour is at least the quota and the tour starts and ends at the root vertex.

The LAQ dataset consists of four graphs\footnote{We constructed smaller graphs because running our algorithms on the entire London road network (which has hundreds of thousands of vertices and edges) was too computationally expensive.} named laqbb, laqid, laqkx and laqws which are each created by only keeping vertices within 3000m of the root vertex.
The root vertex of each graph is given respectively by the following four locations in London: Big Ben (bb), Isle of Dogs (id), King's Cross (kx), and Wembley Stadium (ws).
The prize function of vertices is generated by splitting each edge $(i,j)$ in the graph into two edges $(i, k), (k, j)$ and assigning the length of edge $(i,j)$ to the prize of a newly created vertex $k$.
The prize of $i$ and $j$ is zero.
The cost of $(i, k)$ is the same as $(i, j)$ but the cost of $(k, j)$ is set to zero.
The metric surplus $\zeta(G, c)$ (see \Cref{def:metric-surplus}) of instances in the LAQ dataset is approximately 0.9.
For each of the four graphs, we run our algorithms on five different quotas (1000m, 2000m, 3000m, 4000m, or 5000m), giving us 20 different instances for the London air quality dataset.

\paragraph{TSPLIB} A well-known benchmark dataset used in multiple papers \cite{applegate1995,Berube2009,Fischetti1998}, TSPLIB \cite{Reinelt1991} is a collection of complete graphs.
We chose nine graphs with a range of sizes from the original TSPLIB dataset: eil51\footnote{Code ``eil51'' means there are 51 vertices in the original TSPLIB instance.}, st70, rat195, tsp225, a280, pr439, rat575, gr666, pr1002.
We construct sparse graphs with a metric surplus of zero (meaning every edge is non-metric).
Each vertex is labelled from $1, \ldots, n$ and is assigned an x and y co-ordinate.
The root vertex is labelled $v_1 = 1$.
Let the Euclidean distance between the co-ordinates of vertices $i$ and $j$ be $|| i - j ||_2$.
The prize $p(i)$ of vertex $i$ is defined by three different generations \cite{Berube2009,Fischetti1998}: (i) $p(i) = 1$; (ii) $p(i) = 1 + (7141i + 73) \mod{100}$; and (iii) $p(i) = 1 + \lfloor \frac{99}{\theta} || \rootvertex - i ||_2 \rfloor$ where $\theta = \max_{j \in V(G)} || \rootvertex - j ||_2$.
We set $Q = \alpha \cdot p(V(G))$ where $\alpha \in \{0.05, 0.10, 0.25, 0.50, 0.75 \}$.
To make the graphs sparse, we remove edges from TSPLIB instances with independent and uniform probability until the number of edges is equal to $\kappa n$ where $\kappa \in \{ 5, 10, 15, 20, 25 \}$.

The cost function is called MST with $\zeta(G, c) = 0$ and is defined as follows.
First, assign all edges cost $|| i - j ||_2$.
Next, find the minimum spanning tree $T$ of the sparse graph $G$.
Now, find the shortest path $\cP_{ij}^T$ from vertex $i$ to $j$ using only edges in $E(T)$.
Finally, assign costs such that edges in $T$ are metric and edges not in $T$ are non-metric:
\begin{align*}
    c(i,j) = \begin{cases} 
        \left\lceil || i - j ||_2 \right\rceil & \forall (i, j) \in E(T) \\
        \left\lceil || i - j ||_2 \right\rceil + c(E(\cP_{ij}^T)) & \forall (i, j) \in E(G) \backslash E(T) \\
    \end{cases}
\end{align*}
In summary, there are 675 instances in the modified TSPLIB dataset: five levels of sparsity, three prize generating functions, nine distinct graphs, and five values of $\alpha$.

\subsection{Empirical Heuristic Results} \label{sec:empirical-heuristics}

In this section, we compare SBL-PEC against a baseline (BFS-EC) from the literature.
The BFS-EC baseline initialises the Extension \& Collapse (EC) \cite{DellAmico1998} with the first cycle detected by a Breadth First Search (BFS).
We also run Suurballe's heuristic (SBL) as a stand alone heuristic.
When calculating the GAP, the upper bound (UB) is the total cost of the heuristic solution.
The lower bound (LB) is the cost of the largest lower bound found by running our branch \& cut algorithm in \Cref{sec:branch-cut} with disjoint-paths cost-cover (DPCC) cuts for a maximum of four hours.
On instances where the optimal solution is found within four hours, LB will be the cost of the optimal solution (denoted with a $^\star$ in \Cref{tab:LAQ_compare_heuristics} and \Cref{tab:tspwplib_compare_heuristics}).
For a given instance, each heuristic is compared against the same LB when calculating the GAP.
From \Cref{tab:LAQ_compare_heuristics} and \Cref{tab:tspwplib_compare_heuristics}, we conclude the following:
\begin{itemize}
\item {\it SBL finds low-cost initial tours.} On the LAQ dataset, the \gap\xspace is equal for SBL and SBL-PEC on every instance, implying PEC does not improve the SBL tour because SBL finds a solution that is optimal or close-to-optimal, so little improvement is possible. On TSPLIB, PEC is needed to make the SBL tour prize-feasible.
\item {\it BFS-EC~\cite{DellAmico1998} cannot sufficiently increase the prize.} On both LAQ and TSPLIB datasets, BFS-EC finds less feasible tours than SBL-PEC. On TSPLIB when $G$ is sparse~($\kappa = 5$), BFS-EC finds 8/135 feasible tours compared to 123/135 for SBL-PEC.
\item {\it SBL-PEC is the best heuristic on TSPLIB.} In addition to finding the most feasible tours, when $\kappa \geq 10$, the \gap\xspace of BFS-EC is consistently four times more than the \gap\xspace of SBL-PEC.
\end{itemize}
To conclude our heuristic analysis, SBL is effective as a standalone heuristic on the LAQ dataset, whilst SBL-PEC outperforms all other algorithms on the TSPLIB dataset across varying levels of sparsity.

\begin{table}[t]
    \centering
    \caption{Heuristic comparison on all 4 LAQ instances. ``nan'' means ``not a number'' where we could not calculate the \gap\xspace due to the heuristic finding zero feasible solutions to the four instances. \gap\xspace is marked by $^\star$ if LB is the optimal solution to all four instances found by branch \& cut; else \gap\xspace is unmarked such that LB is the largest lower bound and the \gap\xspace is an overestimate.}
    \label{tab:LAQ_compare_heuristics}
    \begin{tabular}{l
    S[round-mode=places,round-precision=3,scientific-notation=false,table-format=1.3]rc    
    S[round-mode=places,round-precision=3,scientific-notation=false,table-format=1.3]rc
    S[round-mode=places,round-precision=3,scientific-notation=false,table-format=1.3]r}
\toprule
{} & \multicolumn{2}{c}{BFS-EC} && \multicolumn{2}{c}{SBL} && \multicolumn{2}{c}{SBL-PEC}\\\cmidrule{2-3}\cmidrule{5-6}\cmidrule{8-9}
{Quota} & {$\overline{\text{GAP}}$} & {FEAS} && {$\overline{\text{GAP}}$} & {FEAS} && {$\overline{\text{GAP}}$} & {FEAS}  \\
\midrule
1000 & 0.066245{$^\star$} & 3 && 0.000000{$^\star$} & 4 && 0.000000{$^\star$} & 4 \\
2000 & 0.000000{$^\star$} & 1 && 0.039909{$^\star$} & 4 && 0.039909{$^\star$} & 4 \\
3000 & nan & 0 && 0.027181{$^\star$} & 4 && 0.027181{$^\star$} & 4 \\
4000 & nan & 0 && 0.048623{$^\star$} & 4 && 0.048623{$^\star$} & 4 \\
5000 & nan & 0 && 0.116738 & 4 && 0.116738 & 4 \\
\bottomrule
\end{tabular}

\end{table}
\begin{table}[t]
    \centering
    \caption{Heuristic comparison for different sparsity levels~$\kappa$ in the TSPLIB dataset with the MST cost function. Each row corresponds to 135 instances. \gap\xspace is calculated using the largest lower bound (LB), and so the \gap\xspace is an overestimate for each entry.}
    \label{tab:tspwplib_compare_heuristics}
    \begin{tabular}{l
    S[round-mode=places,round-precision=3,scientific-notation=false,table-format=1.3]rc
    S[round-mode=places,round-precision=3,scientific-notation=false,table-format=1.3]rc
    S[round-mode=places,round-precision=3,scientific-notation=false,table-format=1.3]r}
\toprule
{} & \multicolumn{2}{c}{BFS-EC} && \multicolumn{2}{c}{SBL} && \multicolumn{2}{c}{SBL-PEC} \\\cmidrule{2-3}\cmidrule{5-6}\cmidrule{8-9}
{$\kappa$} & {$\overline{\text{GAP}}$} & {FEAS} && {$\overline{\text{GAP}}$} & {FEAS} && {$\overline{\text{GAP}}$} & {FEAS} \\
\midrule
5 & 2.555024 & 8 && 0.097324 & 46 && 1.075363 & 123 \\
10 & 5.430327 & 71 && 0.124323 & 48 && 1.333491 & 135 \\
15 & 8.216562 & 122 && 0.153349 & 51 && 1.420997 & 135 \\
20 & 6.619192 & 120 && 0.183391 & 51 && 1.469251 & 135 \\
25 & 8.124762 & 126 && 0.238618 & 54 && 1.466433 & 135 \\
\bottomrule
\end{tabular}

\end{table}

\newpage

\subsection{Cost-cover Inequalities} \label{sec:cost-cover-results}

To evaluate our disjoint-paths cost-cover (DPCC) cuts, we run our branch \& cut algorithm from \Cref{sec:branch-cut} with DPCC against shortest-path cost-cover (SPCC) cuts.
To find an upper bound $C_U$, we run SBL-PEC.
In addition to TIME and GAP, we record PRE-CUTS:
the number of cost-cover cuts added to the first linear program of the root node of the branch \& cut tree before the solver is started due to the initial upper bound $C_U$.
We found PRE-CUTS to be a more informative metric than total number of cost-cover cuts because the latter can be affected by factors such as the number of times an upper bound is found during the solving process.
From \Cref{tab:LAQ_cost_cover} and \Cref{tab:tspwplib_cost_cover}, we conclude:
\begin{itemize}
\item {\it DPCC adds more \precuts\xspace than SPCC.} Across all quotas on LAQ dataset, DPCC adds at least 300 more \precuts\xspace than SPCC. On TSPLIB for $\alpha=0.05$ and $\alpha = 0.10$ respectively, DPCC adds 61 and 23 more \precuts\xspace than SPCC, resulting in DPCC finding 5 more and 3 more optimal solutions than SPCC respectively. This is empirical evidence of \Cref{prp:cost-cover}.

\item {\it On LAQ, DPCC finds the optimal solution three times faster than SPCC when $Q \leq 4000$.}
Furthermore, the \gap\xspace of DPCC is three times less than SPCC when $Q = 5000$.
Both DPCC and SPCC find 17 out of 20 optimal solutions.

\item {\it Few \precuts~are added when $\alpha = 0.25, 0.50, 0.75$.} 
The result is less than 90 second \avgtime\xspace difference between DPCC and SPCC for $\alpha = 0.25, 0.50, 0.75$.
Few \precuts~are added because larger $\alpha$ means collecting more prize, which will generally incur a greater cost, thus increasing the initial upper bound $C_U^\star$ at the root of the branch \& bound tree.
For all least-cost vertex-disjoint paths~$\cD_1, \cD_2$ from $\rootvertex$ to $t \in V(G)$, if $c(E(\cD_1)) + c(E(\cD_2)) < C_U^\star$, then zero DPCC cuts will be added because the condition in (\ref{eq:disjoint-path-cost-cover}) is never satisfied.
A similar argument can be made for SPCC.
\end{itemize}

\begin{table}[t]
    \small
    \centering
    \caption{Evaluation of our branch \& cut algorithm using disjoint-paths vs shortest-path cost-cover inequalities on the London air quality dataset for five different quotas. There are four instances for each quota.}
    \label{tab:LAQ_cost_cover}
    \begin{tabular}{l
    S[scientific-notation=false,round-mode=places,round-precision=0]
    S[scientific-notation=false,round-mode=places,round-precision=1]
    S[round-mode=places,round-precision=3,scientific-notation=false,table-format=1.3]
    S[scientific-notation=false,drop-zero-decimal=true]
    c
    S[scientific-notation=false,round-mode=places,round-precision=0]
    S[scientific-notation=false,round-mode=places,round-precision=1]
    S[round-mode=places,round-precision=3,scientific-notation=false,table-format=1.3]
    S[scientific-notation=false,drop-zero-decimal=true]
}
    \toprule
    {} & \multicolumn{4}{c}{Disjoint-paths cost-cover cuts} && \multicolumn{4}{c}{Shortest-path cost-cover cuts} \\ \cmidrule{2-5} \cmidrule{7-10}
    {Quota} & {\precuts} & {\avgtime\xspace(s)} & {\gap} & {OPT} && {\precuts} & {\avgtime\xspace(s)} & {\gap} & {OPT} \\
    \midrule
    1000 & 8259.500000 & 26.787274 & 0.000000 & 4 && 7952.250000 & 70.925208 & 0.000000 & 4 \\
    2000 & 8206.750000 & 28.577490 & 0.000000 & 4 && 7840.250000 & 104.258878 & 0.000000 & 4 \\
    3000 & 7967.750000 & 95.886785 & 0.000000 & 4 && 7400.500000 & 3157.767268 & 0.000000 & 4 \\
    4000 & 7405.500000 & 1413.033036 & 0.000000 & 4 && 6788.250000 & 3976.157785 & 0.000000 & 4 \\
    5000 & 6654.000000 & 10870.646244 & 0.041913 & 1 && 5966.500000 & 10981.088769 & 0.151232 & 1 \\
    \bottomrule
\end{tabular}

\end{table}

\begin{table}[t]
    \small
    \centering
    \caption{Evaluation of our branch \& cut algorithm using disjoint-paths vs shortest-path cost-cover inequalities on the TSPLIB dataset for five different values of $\alpha$. Each row corresponds to 135 instances.}
    \label{tab:tspwplib_cost_cover}
    \begin{tabular}{l
    S[scientific-notation=false,round-mode=places,round-precision=0]
    S[scientific-notation=false,round-mode=places,round-precision=1]
    S[round-mode=places,round-precision=3,scientific-notation=false,table-format=1.3]
    S[scientific-notation=false,drop-zero-decimal=true]
    r
    S[scientific-notation=false,round-mode=places,round-precision=0]
    S[scientific-notation=false,round-mode=places,round-precision=1]
    S[round-mode=places,round-precision=3,scientific-notation=false,table-format=1.3]
    S[scientific-notation=false,drop-zero-decimal=true]
}
    \toprule
    {} & \multicolumn{4}{c}{Disjoint-paths cost-cover cuts} && \multicolumn{4}{c}{Shortest-path cost-cover cuts}\\\cmidrule{2-5}\cmidrule{7-10}
    {$\alpha$} & {\precuts} & {\avgtime\xspace(s)} & {\gap} & {OPT} && {\precuts} & {\avgtime\xspace(s)} & {\gap} & {OPT}  \\
    \midrule
    0.05 & 191.666667 & 1922.449267 & 0.045804 & 120 && 131.081481 & 2513.579777 & 0.061733 & 115 \\
    0.10 & 59.392593 & 4993.910444 & 0.349063 & 92 && 36.511111 & 5216.464076 & 0.359163 & 89 \\
    0.25 & 2.222222 & 9030.966884 & 0.768989 & 54 && 0.829630 & 8990.117816 & 0.737785 & 56 \\
    0.50 & 0.000000 & 9272.233008 & 0.826619 & 51 && 0.000000 & 9363.756192 & 0.861708 & 50 \\
    0.75 & 0.000000 & 8531.138686 & 0.669019 & 61 && 0.000000 & 8457.236272 & 0.652404 & 61 \\
    \bottomrule
\end{tabular}

\end{table}

\section{Conclusion}

To summarise, we have designed algorithms for the Prize-collecting Travelling Salesperson Problem on sparse graphs with non-metric cost functions.
\sbl\xspace found optimal or close-to-optimal solutions on the London Air Quality dataset and proved to be a good starting solution for PEC on the TSPLIB dataset.
Our combined SBL-PEC heuristic found a solution with less cost than Extension \& Collapse \cite{DellAmico1998} on every instance across both datasets.
We derived a disjoint-paths cost-cover (DPCC) cut that tightens the lower bound, then proved the size of the set of solutions defined by the DPCC inequality is at most the size of the set of solutions defined by the SPCC inequality.
On the London Air Quality dataset, our DPCC branch \& cut algorithm finds optimal solutions three times faster than SPCC for quotas 1km to 4km, whilst the \gap\xspace is three times less when the quota is 5km.
On the TSPLIB dataset with non-metric costs, DPCC solves more instances to optimality than SPCC.

Our methods can be applied to other family members of TSPs with Profits \cite{Feillet2005,Fischetti1998} and other related routing problems \cite{Debdatta2021,Debdatta2022}.
One example is the Orienteering Problem (OP): maximise the total prize of a tour starting and ending at the root vertex such that the total cost is at most an upper bound $Z \in \mathbb{N}$.
If the least cost of two vertex-disjoint paths between vertices $i,j$ is greater than $Z$, then $i$ and $j$ cannot both be in a feasible solution to OP.
Therefore, the DPCC inequality for \pctsp can be re-written as a valid inequality for OP which is stronger than the SPCC inequality (by \Cref{prp:cost-cover}).

Future work on the sparse, non-metric Prize-collecting Travelling Salesperson Problem could exploit the rich problem structure offered by our motivating example of running routes that minimise air pollution:
dynamic algorithms that optimise a route as air pollution changes over time;
multi-objective algorithms that optimise over multiple pollutants;
and stochastic optimisation algorithms that consider the uncertainty in predictions from the air quality model.



\begin{acknowledgments}
    PO and TD acknowledge support from a UKRI Turing AI Acceleration Fellowship [EP/V02678X/1] and a Turing Impact Award from the Alan Turing Institute.
    PO also acknowledges support from the Department of Computer Science Studentship of the University of Warwick.
    MSR acknowledges support from UKRI EPSRC Research Grant [EP/V044621/1].
    Batch Computing facilities were provided by the Department of Computer Science of the University of Warwick.
    The authors would like to thank Oliver Hamelijnck for providing air quality forecasts of London for the experiments.
    For the purpose of open access, the authors have applied a Creative Commons Attribution (CC-BY) license to any Author Accepted Manuscript version arising from this submission
  \end{acknowledgments}

\bibliography{library}

\newpage

\appendixtitleon
\begin{appendices}
  \newpage
\begin{table*}[t]
    \centering
    \caption{Summary of the assumptions made by the literature on \pctsp. Assumptions include is the cost function metric? Is the graph complete? Is the returned solution a simple cycle? Is the graph directed?}
    \label{tab:literature}
\begin{tabular}[t]{llrrrr}
    \toprule
    Authors & Method & Metric? & Complete? & Simple cycle? & Directed? \\
    \midrule
    Our work & Heuristics + branch \& cut & no & no & yes & no \\
    \citeauthor{Awerbuch1995} \cite{Awerbuch1995} & Approximation algorithm & yes & no & no & no \\
    \citeauthor{Balas1989} \cite{Balas1989,Balas1995,Balas2007} & Polyhedral results & no & yes & yes & yes \\
    \citeauthor{Berube2009} \cite{Berube2009} & Branch \& cut & yes & yes & yes & no \\
    \citeauthor{Chaves2008} \cite{Chaves2008} & Clustering meta-heuristic & no & yes & yes & no \\
    \citeauthor{DellAmico1998} \cite{DellAmico1998,DellAmico1995} & Extension \& Collapse & yes & yes & yes & no \\
    \citeauthor{Fischetti1988} \cite{Fischetti1998} & Lower bounds & no & yes & yes & yes \\
    \citeauthor{Pantuza2022} \cite{Pantuza2022} & Lagrangian relaxation & yes & yes & yes & yes \\ 
    \citeauthor{Pedro2013} \cite{Pedro2013} & Meta-heuristic & no & yes & yes & yes \\
    \bottomrule
\end{tabular}
\end{table*}
\section{Complete graph constructions} \label{sec:constructions}

As is shown in \Cref{tab:literature}, the literature on \pctsp assumes the input graph is complete and/or the cost function is metric.
Given a sparse graph $G$ with a non-metric cost function, one may ask why we cannot simply construct a complete graph $G^\prime$ with metric costs, then run an algorithm from the literature on $G^\prime$?
We give two such graph constructions in this section and explain why they do not work.

\paragraph{Least-cost path construction}
The first construction is as follows: for every pair of vertices $u,v$ in $G$, find the least-cost path $P$ from $u$ to $v$ in $G$, then set the cost of a new edge from $u$ to $v$ in $\gprime$ equal to the cost of $P$.
The problem with such a construction is that a feasible solution to the constructed instance on $G^\prime$ is not guaranteed to be a simple cycle in the original input graph $G$.
Moreover, this construction may not be desirable when $G$ is sparse because the number of edges in $\gprime$ will be far greater than $G$, resulting in more variables to optimise when calling algorithms such as linear programs.

\paragraph{Dummy edge construction}
The second construction takes as input a sparse graph~$G$ and constructs a complete graph $G^\prime$ as follows.
Add a vertices in $G$ to $G^\prime$ and add all edges in $G$ to $G^\prime$ with the same cost.
Now, for all pairs of vertices~$u,v$ in $G$ such that $(u,v)$ is not an edge in $G$, we add a {\em dummy edge}~$(u,v)$ to $G^\prime$ and set the cost of $(u,v)$ to be some very large constant $C$ (e.g. we could take $C = \sum_{(i,j)\in E(G)}~{c(i,j)}$.
The constructed graph~$G^\prime$ is clearly a complete graph and the optimal solution to any instance does not contain dummy edges.
However, the major problem with this construction is that naively running a heuristic from the literature on $G^\prime$ may return a solution that contains a dummy edge, that is, the returned solution is not guaranteed to be feasible for the original instance.
Moreover, the cost function is still non-metric, which naturally rules out any algorithm from the literature which requires a metric cost function.
Finally, the dummy edge construction also increases the computational complexity of algorithms from the literature because the number of edges has been increased from $\cO(n)$ to $\cO(n^2)$.

\section{Proofs of \Cref{sec:problem-definition}}
For completeness, we give proofs of \Cref{lem:metric-edges} and a proof of correctness for the preprocessing algorithm.
\subsection{Number of metric edges}

\lemmetricedges*
\begin{proof}
    Let $F \subseteq E(G)$ be the set of metric edges in $G$.
    Let the sub-graph $G_F$ be defined by vertices $V(G_F) = V(G)$ and edges $E(G_F) = F$.
    To show $|F| \geq n - 1$, it is sufficient to show that $G_F$ is connected.
    By contradiction, suppose $G_F$ is not connected and let $G_1, \ldots, G_k$ be the connected components of $G_F$.
    Since $G$ is connected, there exists a least-cost path $\cP_{uv} = (u, \ldots, x, y, \ldots, v)$ in $G$ from $u \in V(G_i)$ to $v \in V(G_j)$ such that $i \neq j$ visits an edge $(x, y)$ where $x \in V(G_a)$, $y \in V(G_b)$ and $a \neq b$.
    Note that every sub-path of $\cP_{uv}$ is itself a least-cost path.
    Then edge $(x, y)$ defines a least-cost path from $x$ to $y$, so $(x, y)$ is a metric edge and should have been in $F$.
    This implies every two adjacent edges in $\cP_{uv}$ are in the same connected component of $G_F$, so all $u, v \in V(G_F)$ are in the same connected component and $G_F$ is connected.
\end{proof}

\subsection{Biconnected component pre-processing algorithm} \label{sec:preprocessing}
A {\em biconnected component} of $G$ is a maximal sub-graph such that the removal of a single vertex (and all edges incident on that vertex) does not disconnect the sub-graph.
The biconnected components of an undirected graph can be found in time $\cO(n + m)$ \cite{Hopcroft1973}.
Let $B = \{ C_1, \ldots, C_b \}$ be the set of biconnected components of $G$.
Our pre-processing algorithm removes a vertex $r$ from the input graph if there does not exist a $j \in \{ 1, \ldots, b \}$ such that $r$ and $v_1$ are both in $C_j$.
To see why our pre-processing algorithm is correct, note that every vertex $u_i$ in a feasible tour $\tour$ must be biconnected to $v_1$ (otherwise there would exist some vertex in the tour which must be visited more than once).
A trivial example of a vertex not in the same biconnected component as $v_1$ is a leaf vertex with degree one.

We give a complete proof of correctness for our pre-processing algorithm.
Given a subset of vertices $S \subseteq V(G)$, the {\em induced sub-graph} $G[S]$ contains all vertices in $S$ and contains all edges in $E(G)$ that have both endpoints in $S$.
Formally, a {\em biconnected component} of $G$ is a maximal sub-graph such that the removal of a single vertex (and all edges incident on that vertex) does not disconnect the sub-graph.
We define $C_1, \ldots, C_b \subseteq V(G)$ to be vertex sets such that $G[C_1], \ldots, G[C_b]$ are the biconnected components of $G$.

\begin{lemma} \label{lem:biconnected-intersection}
    Let $G[C_1], G[C_2]$ be any two maximal biconnected components of $G$.
    Then $|C_1 \cap C_2| \leq 1$.
\end{lemma}
\begin{proof}
    Suppose that $|C_1 \cap C_2| > 1$ and let $C^\prime = C_1 \cup C_2$.
    Our claim is that $G[C^\prime]$ is a biconnected component.
    To prove this claim, take two vertices $i,j \in C^\prime$.
    If $i,j \in C_1$ (or if $i,j \in C_2$), then $i$ and $j$ are already in the same biconnected component.
    If $i \in C_1 \backslash (C_1 \cap C_2)$ and $j \in C_2  \backslash (C_1 \cap C_2)$, then remove a vertex $k_1 \in C_1 \cap C_2$.
    This removal does not disconnect the sub-graph $G[C^\prime]$ because there exists a path from $i$ to $j$ via some distinct $k_2 \in C_1 \cap C_2$ where $k_2 \neq k_1$ (since $|C_1 \cap C_2| > 1$).
    $G[C^\prime]$ is therefore a larger biconnected than $G[C_1]$ and $G[C_2]$, so $G[C_1], G[C_2]$ are not {\em maximal} biconnected component components.
\end{proof}

\begin{lemma} \label{lem:bcc-tour}
    Assume $G$ is connected.
    Let $G[C_1], G[C_2]$ be any two maximal biconnected components of $G$ such that $C_1 \neq C_2$.
    Let tour $\tour$ be a feasible solution to an instance of Pc-TSP.
    If $\tour$ visits at least two vertices excluding $v_1$ in $C_1$, i.e. $|(V(\tour) \cap C_1) \backslash \{ v_1 \}| \geq 2$, then $\tour$ does not visit any vertices in $C_2 \backslash (C_1 \cap C_2)$.
\end{lemma}
\begin{proof}
    From \Cref{lem:biconnected-intersection}, we have two cases: (i) $C_1 \cap C_2$ contains exactly one vertex $r$; or (ii) $C_1 \cap C_2 = \emptyset$.
    For case (i), if $\tour$ visits a vertex $j \in C_2 \backslash (C_1 \cap C_2)$, then $\tour$ must visit vertex $r$ twice (once on the way to $j$, and once on the way back) which is contradiction to the definition of a tour.
    For case (ii), there exists exactly one biconnected component\footnote{We can construct an undirected graph $H$ by representing each biconnected components as a vertex in $H$. Note $H$ will be acyclic. If $G$ is connected, then $H$ is a connected tree.} $G[C_3]$ such that $C_3 \neq C_1 \neq C_2$ and $C_1 \cap C_3 \neq \emptyset$.
    Any tour that visits a vertex in $C_2$ must also visit a vertex in $j \in C_3 \backslash (C_1 \cap C_3)$.
    By the same logic as case (i), no tour can visit $j$, therefore no tour can visit $C_2$.
\end{proof}

Applying \Cref{lem:bcc-tour} to every biconnected component that does not contain the root leads to our main pre-processing result:

\begin{corollary}
    Let $R = \bigcup \{ C_k | v_1 \in C_k, k \in \{ 1, \ldots, b \} \}$ be the set of vertices in the same biconnected component as the root vertex.
    There does not exist a feasible solution $\tour$ that visits a vertex in $V(G) \backslash R$.
    Vertices in $V(G) \backslash R$ (and their adjacent edges) can safely be removed from the input graph $G$ without removing any feasible solutions.
\end{corollary}

\section{Proofs of \Cref{sec:heuristics}} \label{sec:heuristics-proof}

\feasibleheuristics*
\begin{proof}
    By contradiction, assume $A$ is guaranteed to find a feasible solution for every instance of \pctsp in polynomial-time.
    Let $\cI = (G, c, p, Q, \rootvertex)$ be an instance such that $Q = p(V(G))$.
    The only feasible solutions to $\cI$ are Hamiltonian cycles of $G$.
    Then $A$ is guaranteed to find a Hamiltonian cycle in polynomial-time for any simple, undirected graph $G$.
    Since finding a Hamiltonian cycle \nphard, this implies ${\cal P} = \NP$, a contradiction to our assumption.
\end{proof}

\section{Proofs of \Cref{sec:cost-cover}}
\label{sec:cost-cover-proofs}

\disjoint*
\begin{proof}
    Split $\tour$ into two vertex-disjoint paths~$\cD_1^\prime, \cD_2^\prime$ starting at $i$ and ending at $j$.
    If $c(E(\cD_1^\prime)) + c(E(\cD_2^\prime)) < c(E(\cD_1))+ c(E(\cD_2))$, then $\cD_1, \cD_2$ are not the least-cost vertex-disjoint paths between $i$ and $j$.
    If $c(E(\cD_1^\prime)) + c(E(\cD_2^\prime)) > c(E(\cD_1))+ c(E(\cD_2))$, then there exists a tour $\tour^\prime$ containing $i$ and $j$ with less cost than $\tour$, because we can construct $\tour^\prime$ by reversing the order of $\cD_2$ and appending it to $\cD_1$.
    So $\cD_1^\prime, \cD_2^\prime$ are least-cost vertex-disjoint paths and the minimum cost of the tour is $c(E(\tour)) = c(E(\cD_1^\prime)) + c(E(\cD_2^\prime)) = c(E(\cD_1))+ c(E(\cD_2))$.
\end{proof}

\section{Details of branch \& cut algorithm}

\subsection{Separation Algorithms} \label{sec:separation}

In a branch \& cut algorithm, a separation algorithm identifies violated inequalities.
We give polynomial-time separation algorithms for the sub-tour elimination constraints (SECs).

\begin{figure}[t]
    \centering
    \includegraphics[width=0.85\linewidth]{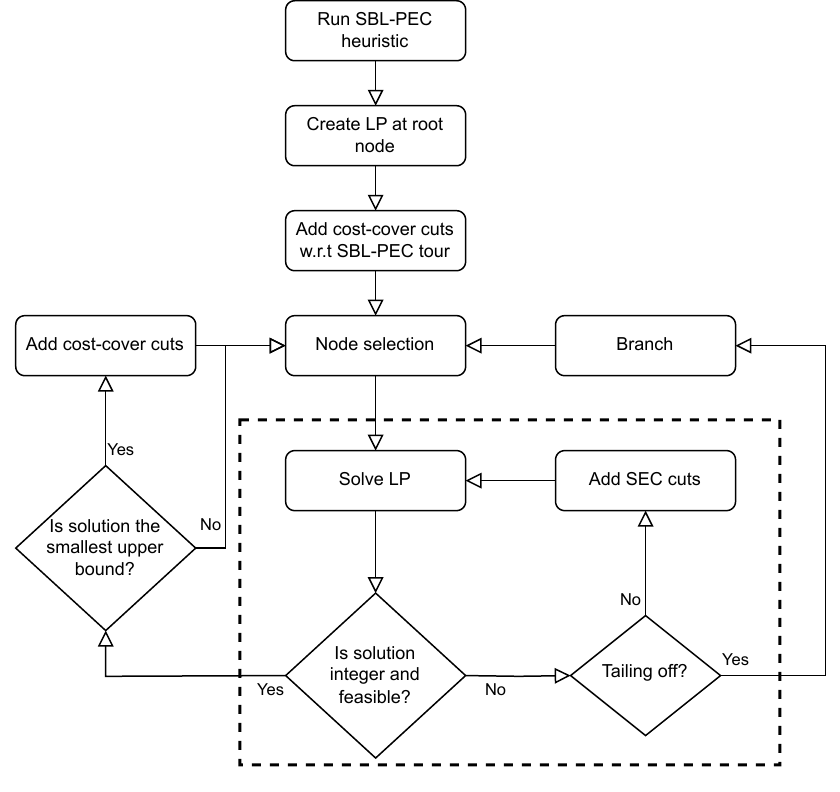}
    \caption{A flow diagram of our branch \& cut algorithm. The dashed square box denotes operations that are repeated within a single node of the branch \& bound tree.}
    \label{fig:branch-and-cut}
\end{figure}

\subsubsection{Sub-tour Elimination Constraints} \label{sec:sec}
\begin{definition}
    The {\em support graph} $\Gstar$ of the solution $(x^\star, y^\star)$ to linear program $LP$ is defined by the vertex set $V(\Gstar) = \{ i \in V(G) : y_i^\star > 0 \}$ and edge set $E(\Gstar) = \{ (i,j) \in E(G) : x_{ij}^\star > 0 \}$.
\end{definition}
Constraints (\ref{con:sub-tour-elimination}) are the SECs.
Since there are an exponential number of SECs, we add violated SECs as a cutting plane.
Given the support graph $\Gstar$ of a linear program $LP$, we run the separation algorithm on $\Gstar$ to check if a set $S \subseteq V(\Gstar)$ violates (\ref{con:sub-tour-elimination}).
First, suppose the support graph is not connected.
Let $\Gstar_1, \ldots, \Gstar_k$ be the $k$ connected sub-graphs of $\Gstar$.
For every sub-graph $\Gstar_l$ where $\rootvertex \notin V(\Gstar_l)$ and $l \in \{ 1,\ldots,k\}$, apply (\ref{con:sub-tour-elimination}) to each vertex $i \in V(\Gstar_l)$ by setting $S = V(\Gstar_l)$.
Now suppose the support graph is connected.
Treat the value of the edge variables~$\bm x$ as the capacity of an edge.
For each $i \in V(\Gstar)$, find a minimum capacity $(\rootvertex, i)$-cut in $\Gstar$ with a min-cut max-flow algorithm in $\cO(|V(\Gstar)|^3)$ \cite{Goldberg1988}.
Let $(S_i, V(\Gstar) \backslash S_i)$ be the minimum capacity cut, where $\rootvertex \in V(\Gstar) \backslash S_i$.
If \cref{con:sub-tour-elimination} is violated for $S_i$, add it to the set of constraints in the linear program.

\subsection{Variable Branching} \label{sec:branching-strategy}
\begin{table*}[t]
    \centering
    \caption{Statistics for the LAQ dataset. The original input graph is $G$ and the pre-processed graph is $H$. The column $\frac{p(V(H))}{p(V(G))}$ denotes the total prize of $H$ as a ratio of $G$. Column $D(G)$ is defined by (\ref{eq:disjoint-prize}).}
    \label{tab:LAQ_dataset}
    \begin{tabular}{l
    S[round-mode=none,group-separator = {,},group-minimum-digits = 4,scientific-notation=false,table-format=5.0]
    S[round-mode=none,group-separator = {,},group-minimum-digits = 4,scientific-notation=false,table-format=5.0]
    S[round-mode=none,table-format=6.0,group-separator = {,},group-minimum-digits = 4,scientific-notation=false]
    S[round-mode=none,table-format=5.0,group-separator = {,},group-minimum-digits = 4,scientific-notation=false]
    S[round-mode=none,table-format=5.0,group-separator = {,},group-minimum-digits = 4,scientific-notation=false]
    S[round-mode=none,table-format=6.0,group-separator = {,},group-minimum-digits = 4,scientific-notation=false]
    S[round-mode=places,round-precision=3,scientific-notation=false,table-format=1.3]
    S[round-mode=places,round-precision=3,scientific-notation=false,table-format=1.3]
    S[round-mode=places,round-precision=3,scientific-notation=false,table-format=1.3]}
\toprule
{Graph name} & {$|V(G)|$} & {$|E(G)|$} & {$p(V(G))$} & {$|V(H)|$} & {$|E(H)|$} & {$p(V(H))$} & {$\frac{p(V(H))}{p(V(G))}$} & {$\zeta(G, c)$} & {$D(G)$} \\
\midrule
laqbb & 14530 & 16791 & 595242 & 11575 & 13786 & 501059 & 0.841774 & 0.910698 & 0.020703  \\
laqid & 11373 & 12541 & 481922 & 6939 & 8046 & 342243 & 0.710163 & 0.857143 & 0.040903 \\
laqkx & 13431 & 15427 & 579233 & 10396 & 12328 & 474423 & 0.819054 &  0.909364 & 0.022428 \\
laqws & 6770 & 7519 & 365947 & 4223 & 4912 & 256929 & 0.702093 & 0.898667 & 0.036478 \\
\bottomrule
\end{tabular}

\end{table*}

\begin{table*}[t]
    \centering
    \caption{Statistics for the TSPLIB dataset aggregated by the sparsity level $\kappa$ and the cost function (EUC or MST). The original input graph is $G$ and the pre-processed graph is $H$. The column $\text{AVG}\left(\frac{p(V(H))}{p(V(G))}\right)$ denotes the average total prize of $H$ as a ratio of $G$. The column $\text{AVG}(D(G))$ is an average of $D(G)$ as defined by (\ref{eq:disjoint-prize}).}
    \label{tab:tspwplib_dataset}
    \begin{tabular}{l
    S[round-mode=places,round-precision=3,scientific-notation=false,table-format=1.3]
    S[round-mode=places,round-precision=3,scientific-notation=false,table-format=1.3]
    c
    S[round-mode=places,round-precision=3,scientific-notation=false,table-format=1.3]
    S[round-mode=places,round-precision=3,scientific-notation=false,table-format=1.3]}
\toprule
& \multicolumn{2}{c}{EUC} && \multicolumn{2}{c}{MST} \\\cmidrule{2-3}\cmidrule{5-6}
{$\kappa$} & {\text{AVG}$(D(G))$} & {\text{AVG}$\left(\frac{p(V(H))}{p(V(G))}\right)$} && {\text{AVG}$(D(G))$} & {\text{AVG}$\left(\frac{p(V(H))}{p(V(G))}\right)$} \\
\midrule
5 & 0.061698 & 0.911594 && 0.138146 & 0.911594 \\
10 & 0.050040 & 0.993222 && 0.140432 & 0.993222 \\
15 & 0.044225 & 0.998673 && 0.164034 & 0.998673 \\
20 & 0.041317 & 1.000000 && 0.189058 & 1.000000 \\
25 & 0.037304 & 1.000000 && 0.184497 & 1.000000 \\
\bottomrule
\end{tabular}
\end{table*}

When a linear program $LP$ at node $N$ in the branch \& bound tree is solved,
we may choose to branch on a variable $x_{ij}$ that is fractional in $LP$.
Branching on a variable means creating two new child nodes $N^+, N^-$ of $N$ with linear programs $LP^+$ and $LP^-$ respectively.
$LP^+$ is defined by inheriting all variables and constraints in $LP$ and then rounding variable $x_{ij}$ up to the nearest integer ($x_{ij} = 1$).
$LP^-$ is defined similarly by rounding $x_{ij}$ down ($x_{ij} = 0$).
Choosing which fractional variable to branch on has been the subject of extensive research \cite{Achterberg2005}.
In their branch \& cut algorithm for \pctsp, \citeauthor{Berube2009} \cite{Berube2009} branch on the most fractional variable.
However, more advanced branching strategies can yield smaller branch \& bound trees, reducing the total computational time for solving a problem to optimality.
Our branch \& cut algorithm uses strong branching \cite{applegate1995} at the top node of the branch \& bound tree, then uses reliability branching \cite{Achterberg2005} on all subsequent nodes in the tree.


We make one final addition to our branch \& cut algorithm to prevent the phenomena of {\em tailing off}.
This is where adding cuts to consecutive relaxed linear programs from the same node does little to improve the lower bound.
To test if a node in the branch \& bound tree is tailing off, we check if $\text{GAP}_t - \text{GAP}_{t - \tau} \leq \gamma$ where 
$\text{GAP}_t$ is the GAP between the lower bound and upper bound after $t$ iterations of adding cuts,
$\tau \in \mathbb{N}$ is a parameter controlling the maximum number of iterations cuts can be added to a linear program without showing improvement,
and $\gamma \in \mathbb{R}$ is the LP gap improvement threshold.
If we find a node is tailing off, we branch on a variable, instead of waiting for more cuts to be added.
After some experimentation (see \Cref{sec:branching-and-tailing-off}), we set $\tau = 5$ and $\gamma=0.001$.
To summarise, if after five iterations of adding cuts to a linear program the GAP has improved by at most 0.001, we branch on a variable instead of adding more cuts.
\Cref{fig:branch-and-cut} shows a visualisation of our branch \& cut algorithm.

\section{Datasets and preprocessing} \label{sec:datasets-preprocessing}

Before outlining the complementary experiments, we give more details about the LAQ and TSPLIB datasets.
Moreover, for the TSPLIB dataset, we introduce a {\em metric} cost function called EUC.
It is defined simply as the Euclidean distance between the coordinates of two vertices~$i,j \in V(G)$ in the TSPLIB dataset: $c(i,j) = ||i - j||_2$.

\Cref{tab:LAQ_dataset} shows a summary of instances in the LAQ dataset and \Cref{tab:tspwplib_dataset} shows a summary of instances in the TSPLIB dataset.
The table includes a graph property $D: G \rightarrow [0, 1]$ which is defined to be the largest prize of a pair of least-cost vertex-disjoint paths $\cD_1, \cD_2$ from  $\rootvertex$ to $t \in V(G)$ as a ratio of the total prize of the graph:
\begin{align} \label{eq:disjoint-prize}
    D(G) = \frac{1}{p(V(G))}\max_{t \in V(G)} \{ p(V(\cD_1)) + p(V(\cD_2)) - p(t) \} 
\end{align}

\subsection{Pre-processing} \label{sec:pre-processing-results}

The biconnected component pre-processing algorithm described in \Cref{sec:preprocessing} removes more vertices (and more prize) from the LAQ input graph than on the TSPLIB input graph because the LAQ graphs are more sparse than TSPLIB.
From \Cref{tab:tspwplib_dataset}, we see that on TSPLIB the average prize removed from the graph during pre-processing increases as the graph becomes more sparse ($\kappa$ decreases).
Even when $\kappa = 5$ on TSPLIB, less than 10\% of the prize is removed from the input graph during pre-processing.
In contrast on the LAQ dataset in \Cref{tab:LAQ_dataset}, our pre-processing algorithm removes more than 25\% of the prize on laqid and laqws, and removes 15.8\% and 18.1\% on laqbb and laqkx.
To summarise, our pre-processing algorithm is effective at reducing the input size when the graph is sparse.

\section{Additional heuristic results} \label{sec:additional-heuristic-results}


\begin{table*}
    \centering
    \label{tab:tspwplib_compare_heuristics_extra}
    \caption{The \gap~and FEAS~for five heuristics on the TSPLIB dataset across both the EUC and MST cost functions and across different sparsity levels~$\kappa$. Each row corresponds to 135 instances. \gap\xspace is calculated using the largest lower bound (LB) and so the \gap\xspace is an overestimate for each entry.}
    \begin{tabular}{ll
    S[round-mode=places,round-precision=3,scientific-notation=false,table-format=1.3]rc
    S[round-mode=places,round-precision=3,scientific-notation=false,table-format=1.3]rc
    S[round-mode=places,round-precision=3,scientific-notation=false,table-format=1.3]rc
    S[round-mode=places,round-precision=3,scientific-notation=false,table-format=1.3]rc
    S[round-mode=places,round-precision=3,scientific-notation=false,table-format=1.3]r
}
\toprule
{} & & \multicolumn{2}{c}{BFS-EC} && \multicolumn{2}{c}{BFS-PEC} && \multicolumn{2}{c}{SBL-EC}  && \multicolumn{2}{c}{SBL} && \multicolumn{2}{c}{SBL-PEC} \\\cmidrule{3-4}\cmidrule{6-7}\cmidrule{9-10}\cmidrule{12-13}\cmidrule{15-16}
{Cost} & {$\kappa$} & {$\overline{\text{GAP}}$} & {FEAS} && {$\overline{\text{GAP}}$} & {FEAS} && {$\overline{\text{GAP}}$} & {FEAS} && {$\overline{\text{GAP}}$} & {FEAS} && {$\overline{\text{GAP}}$} & {FEAS} \\
\midrule
EUC & 5 & 0.506064 & 11 && 1.398531 & 42 && 0.674543 & 124 && 0.040846 & 18 && 0.579893 & 125 \\
 & 10 & 2.464662 & 79 && 2.535878 & 93  && 0.750023 & 135 && 0.161855 & 15 && 0.735019 & 135 \\
 & 15 & 3.357539 & 117 && 3.372375 & 121 && 0.803059 & 135 && 0.192899 & 12 && 0.797125 & 135 \\
 & 20 & 3.035725 & 106 && 3.593715 & 120 && 0.905661 & 135 && 0.242203 & 11 && 0.798989 & 135 \\
 & 25 & 3.170166 & 107 && 2.918703 & 121 && 0.815840 & 135 && 0.460362 & 8 && 0.776364 & 135 \\
MST & 5 & 2.555024 & 8 && 3.325319 & 62  && 1.642203 & 122 && 0.097324 & 46 && 1.075363 & 123 \\
 & 10 & 5.430327 & 71 && 6.108636 & 97 && 1.789664 & 135 && 0.124323 & 48 && 1.333491 & 135 \\
 & 15 & 8.216562 & 122 && 6.617239 & 125 && 1.862620 & 135 && 0.153349 & 51 && 1.420997 & 135 \\
 & 20 & 6.619192 & 120 && 6.141395 & 133 && 1.800211 & 135 && 0.183391 & 51 && 1.469251 & 135 \\
 & 25 & 8.124762 & 126 && 5.622649 & 133 && 1.825296 & 135 && 0.238618 & 54 && 1.466433 & 135 \\
\bottomrule
\end{tabular}

\end{table*}

\begin{table}[t]
    \centering
    \caption{For a given $\alpha$, each row reports the number of feasible solutions (FEAS) found by a heuristic on the TSPLIB dataset across both the EUC and MST cost functions. The largest possible value of FEAS is 270.}
    \label{tab:tsplib-alpha}
    \begin{tabular}{lrrrr}
        \toprule
        $\alpha$ & BFS-EC & BFS-PEC & \xspace\xspace\xspace\xspace SBL & SBL-PEC \\
        \midrule
        0.05 & 204 & 270 & 181 & 270 \\
        0.10 & 194 & 270 & 107 & 270 \\
        0.25 & 187 & 270 & 26 & 270 \\
        0.50 & 160 & 270 & 0 & 270 \\
        0.75 & 122 & 246 & 0 & 248 \\
        \bottomrule
    \end{tabular}
\end{table}

\Cref{tab:tspwplib_compare_heuristics_extra} shows heuristics results in the same manor as the TSPLIB results of \Cref{tab:tspwplib_compare_heuristics} but with the extra cost function EUC and with two addtional heuristics called BFS-PEC and SBL-EC.
After generating an initial tour with BFS, the BFS-PEC heuristic increases the prize with the path extension heuristic, then decreases the cost with the path collapse heuristic. 
After generating an initial tour with SBL, SBL-EC increases the prize with \texttt{Extension}, then decreasing the cost with \texttt{Collapse} \cite{DellAmico1998}.

Both BFS-PEC and SBL-PEC find feasible solutions to every instance of TSPLIB when $\kappa \geq 10$ and $\alpha \leq 0.50$.
BFS-PEC and SBL-PEC do not find feasible solutions to every instance when the graph is sparse ($\kappa = 5$) and the quota is large ($\alpha = 0.75$): BFS-PEC finds 246 out of 270 feasible solutions and SBL-PEC finds 248 out of 270 across both cost functions.
The \gap\xspace of SBL-PEC is smaller than BFS-PEC across every $\kappa$ for both cost functions, but the difference in the \gap\xspace between the two algorithms is more noticeable on the MST cost function.
When the cost function is non-metric (MST), the difference in the \gap\xspace between BFS-PEC and SBL-PEC is greater than 0.33 for all $\kappa$; and when the cost function is metric (EUC), the difference in the \gap\xspace between BFS-PEC and SBL-PEC is less than 0.11 for all $\kappa$.
This suggests {\em initialising PEC with SBL is particularly effective on non-metric cost functions}.

The GAP of SBL-EC is greater than BFS-PEC and SBL-PEC across both cost functions and all values of $\kappa$.
Moreover, SBL-EC finds less feasible solutions than BFS-PEC and SBL-PEC.
Compared to BFS-EC, SBL-EC finds more feasible solutions, suggesting it is better to initialise EC with SBL than BFS: this is the same conclusion we came to for initialising PEC.
The $\overline{\text{GAP}}$ of SBL-EC is generally greater than BFS-EC on the EUC cost function, but less than BFS-EC on MST cost function (however these figures are not directly comparable because SBL-EC finds more feasible solutions).

Whilst SBL is highly effective on the LAQ dataset, \Cref{tab:tsplib-alpha} shows SBL is less likely to find a feasible solution as $\alpha$ increases on the TSBLIB dataset.
This is unsurprising because a pair of least-cost vertex-disjoint paths between $\rootvertex$ and $t\in V(G)$ is unlikely to collect a lot of prize, thus if $\alpha$ is large then SBL will not collect enough prize to meet the quota.
More specifically, if $D(G) > \alpha$ then SBL will not find a feasible solution.
For example, \Cref{tab:tspwplib_dataset} shows that $\text{AVG}(D(G)) = 0.062$ on the EUC cost function when $\kappa = 5$, so if $\alpha = 0.75$ then SBL will not find a feasible solution.
One reason SBL is highly effective on the LAQ dataset is that $\alpha$ is small: for example, a quota of 5000 on the laqbb graph gives $\alpha \approx 0.01$ which is less than $D(G) = 0.021$.
To conclude our analysis of heuristics, SBL is effective as a standalone heuristic on the LAQ dataset, whilst SBL-PEC outperforms all other algorithms on the TSPLIB dataset across two cost functions and varying levels of sparsity.

\section{Additional cost-cover results}
\begin{table}[t]
    \centering
    \caption{Evaluation of our branch \& cut algorithm using no cost-cover (NoCC) inequalities on the London air quality dataset for five different quotas. There are four instances for each quota.}
    \label{tab:LAQ_cost_cover_nocc}
    \begin{tabular}{l
    S[scientific-notation=false,round-mode=places,round-precision=0]
    S[scientific-notation=false,round-mode=places,round-precision=1]
    S[round-mode=places,round-precision=3,scientific-notation=false,table-format=1.3]
    S[scientific-notation=false,drop-zero-decimal=true]
}
    \toprule
    {} & \multicolumn{4}{c}{No cost-cover cuts} \\\cmidrule{2-5}
    {Quota} & {\precuts} & {\avgtime\xspace(s)} & {\gap} & {OPT} \\
    \midrule
    1000 & 0.000000 & 3782.760003 & 0.544130 & 3 \\
    2000 & 0.000000 & 6594.293784 & 0.348440 & 3 \\
    3000 & 0.000000 & 12576.040385 & 0.336305 & 1 \\
    4000 & 0.000000 & 14432.305469 & 0.268663 & 0 \\
    5000 & 0.000000 & 14434.357712 & 0.324544 & 0 \\
    \bottomrule
\end{tabular}

\end{table}

\begin{table}[t]
    \centering
    \caption{Evaluation of our branch \& cut algorithm using no cost-cover (NoCC) inequalities on the TSPLIB dataset for five different values of $\alpha$ and two different cost functions. Each row corresponds to 135 instances.}
    \label{tab:tspwplib_cost_cover_nocc}
    \begin{tabular}{ll
    S[scientific-notation=false,round-mode=places,round-precision=0]
    S[scientific-notation=false,round-mode=places,round-precision=1]
    S[round-mode=places,round-precision=3,scientific-notation=false,table-format=1.3]
    S[scientific-notation=false,drop-zero-decimal=true]
}
    \toprule
    {} & {} & \multicolumn{4}{c}{No cost-cover cuts}\\\cmidrule{3-6}
    {Cost} & {$\alpha$} & {\precuts} & {\avgtime\xspace(s)} & {\gap} & {OPT} \\
    \midrule
    {EUC} & 0.05 & 0.000000 & 2566.818535 & 0.070621 & 118 \\
    & 0.10 & 0.000000 & 3883.357246 & 0.132874 & 104 \\
    & 0.25 & 0.000000 & 4606.162898 & 0.148750 & 101 \\
    & 0.50 & 0.000000 & 5393.606414 & 0.122540 & 95 \\
    & 0.75 & 0.000000 & 5032.733609 & 0.088334 & 94 \\
    {MST} & 0.05 & 0.000000 & 3449.748495 & 0.079908 & 111 \\
    & 0.10 & 0.000000 & 5604.624505 & 0.373282 & 87 \\
    & 0.25 & 0.000000 & 9048.364034 & 0.742481 & 54 \\
    & 0.50 & 0.000000 & 9285.303174 & 0.858871 & 51 \\
    & 0.75 & 0.000000 & 8507.030370 & 0.648810 & 60 \\
    \bottomrule
\end{tabular}

\end{table}
\begin{table*}[t]
    \centering
    \caption{Evaluation of our branch \& cut algorithm using disjoint-paths vs shortest-path cost-cover inequalities on the TSPLIB dataset with EUC cost function for five different values of $\alpha$. Each row corresponds to 135 instances.}
    \label{tab:tspwplib_cost_cover_euc}
    \begin{tabular}{l
    S[scientific-notation=false,round-mode=places,round-precision=0]
    S[scientific-notation=false,round-mode=places,round-precision=1]
    S[round-mode=places,round-precision=3,scientific-notation=false,table-format=1.3]
    S[scientific-notation=false,drop-zero-decimal=true]
    r
    S[scientific-notation=false,round-mode=places,round-precision=0]
    S[scientific-notation=false,round-mode=places,round-precision=1]
    S[round-mode=places,round-precision=3,scientific-notation=false,table-format=1.3]
    S[scientific-notation=false,drop-zero-decimal=true]
}
    \toprule
    {} & \multicolumn{4}{c}{Disjoint-paths cost-cover cuts} && \multicolumn{4}{c}{Shortest-path cost-cover cuts}\\\cmidrule{2-5}\cmidrule{7-10}
    {$\alpha$} & {\precuts} & {\avgtime\xspace(s)} & {\gap} & {OPT} && {\precuts} & {\avgtime\xspace(s)} & {\gap} & {OPT}  \\
    \midrule
    0.05 & 56.748148 & 2239.787574 & 0.075131 & 120 && 50.400000 & 2238.865383 & 0.084406 & 120 \\
    0.10 & 9.022222 & 3846.827889 & 0.138628 & 103 && 8.148148 & 3782.394973 & 0.140535 & 104 \\
    0.25 & 0.222222 & 4630.079474 & 0.130462 & 101 && 0.207407 & 4715.326725 & 0.137173 & 101 \\
    0.50 & 0.000000 & 5464.072675 & 0.125017 & 93 && 0.000000 & 5265.654930 & 0.113322 & 96 \\
    0.75 & 0.000000 & 5167.207855 & 0.056486 & 96 && 0.000000 & 4940.336448 & 0.090504 & 96 \\
     \bottomrule
\end{tabular}

\end{table*}

In this section, we conduct two additional experiments.
Firstly, we ask: does running a branch \& cut algorithm with cost-cover inequalities actually make a different?
We find an affirmative answer by running our branch \& cut algorithm with no cost-cover (NoCC) inequalities and compare it against DPCC and SCPCC.
Second, we analyse DPCC and SPCC on a {\em metric} cost function (EUC).

\subsection{No cost cover inequalities} \label{sec:no-cost-cover}
\Cref{tab:LAQ_cost_cover_nocc} shows that, on the LAQ dataset, adding no cost cover inequalities (NoCC) vastly {\em underperforms} DPCC and SPCC from \Cref{tab:LAQ_cost_cover}.
NoCC finds just 7 out of 20 optimal solutions, whereas DPCC and SPCC both find 17 out of 20.
We conclude that adding cost cover inequalities is highly advantageous on the LAQ dataset.

On TSPLIB with MST costs when $\alpha=0.05$ and $\alpha = 0.10$, NoCC adds zero cost-cover cuts, so DPCC finds 9 more and 5 more optimal solutions than NoCC for $\alpha=0.05$ and $\alpha = 0.10$ respectively.
The \avgtime\xspace of DPCC is less than NoCC when $\alpha=0.05$ and $\alpha = 0.10$: DPCC terminates 1,530 and 610 seconds faster respectively than NoCC.

However, when $\alpha \in \{ 0.25, 0.50, 0.75 \}$ for MST costs, the number of \precuts\xspace is small or zero for both DPCC and SPCC.
Indeed when DPCC and SPCC add zero cost-cover cuts in total, then DPCC, SPCC, and NoCC are nearly equivalent branch \& cut algorithms (minus the added computational time needed to run the cost-cover separation algorithm).
This results in less than 90 second \avgtime\xspace difference between DPCC, SPCC, and NoCC for all $\alpha \in \{ 0.25, 0.50, 0.75 \}$.
Small variations in the \gap\xspace and OPT further support that there is little difference on MST costs between DPCC, SPCC, and NoCC when $\alpha \in \{ 0.25, 0.50, 0.75 \}$: SPCC has the smallest \gap\xspace by 0.004 when $\alpha = 0.25$; DPCC has the smallest \gap\xspace by 0.032 when $\alpha = 0.50$; and NoCC has the smallest \gap\xspace by 0.003 when $\alpha = 0.75$.

\subsection{Metric costs}

The EUC cost function on the TSPLIB dataset, as defined in \Cref{sec:datasets-preprocessing}, is a metric cost function (see \Cref{def:metric}).
The cost-cover results for EUC on TSPLIB are shown in \Cref{tab:tspwplib_cost_cover_euc}.
For each $\alpha$, less \precuts\xspace are added by DPCC on EUC compared to when the cost function is MST.
For example, when $\alpha = 0.05$, the number of \precuts\xspace added by DPCC is 56.7 on EUC compared to 192 on MST.
Moreover, for all values of $\alpha$, the difference in the number of \precuts\xspace between DPCC and SPCC on the EUC cost function is smaller than on the difference MST cost function.
As a result, conclude that for the EUC cost function, small variations across in the \avgtime\xspace, \gap\xspace and OPT show there is little difference between DPCC, SPCC, and NoCC.

\section{Varying $\alpha$ on LAQ dataset}

Our results show a discrepancy between the LAQ and TSPLIB datasets.
For completeness, we run the same experiments as \Cref{sec:results} on the LAQ dataset for both our heuristics and our branch \& cut algorithm, but instead of setting the quota to be $Q \in \{ 1000, 2000, 3000, 4000, 5000 \}$, we set the quota $Q = \alpha \cdot p(V(G))$ in the same manor as our experiments on TSPLIB with $\alpha \in \{ 0.05, 0.10, 0.25, 0.50, 0.75 \}$.
The experimental results for heuristics are summarised in \Cref{tab:LAQ_heuristics_alpha} and for different cost-cover cuts in \Cref{tab:LAQ_cost_cover_alpha}.

First, as we noted in \Cref{sec:pre-processing-results}, our biconnected component pre-processing algorithm described in \Cref{sec:problem-definition} removes more vertices (and more prize) from the LAQ input graph than on the TSPLIB input graph because the LAQ graphs are more sparse.
From \Cref{tab:tspwplib_dataset}, we see that on TSPLIB the average prize removed from the graph during pre-processing is less than $10\%$ across all sparsity levels $\kappa$.
Contrast this to the LAQ dataset in \Cref{tab:LAQ_dataset}, where on two LAQ instances (laqid and laqws), our pre-processing algorithm removes more than 25\% of the prize in the original input graph.
This means when $\alpha = 0.75$ for laqid and laqws, the total prize of the pre-processed graph is not greater than the quota, which trivially shows the original instance does not have a feasible solution.
On laqbb and laqkx, the pre-processing algorithm removes 15.8\% and 18.1\% of the total prize of the original graph.
Whilst in theory the two graphs contains sufficient prize for a quota set with $\alpha=0.75$, the bottom row of \Cref{tab:LAQ_cost_cover_alpha} shows our branch \& cut algorithm quickly determines there are no feasible solutions to the two instances.
Moreover, when $\alpha = 0.50$, neither our heuristics nor our branch \& cut algorithm found a feasible solution to any of the four instances.
The branch \& cut algorithm was able to obtain a lower bound, but not able to prove infeasibility within the four hour time limit, and it remains open whether there exists a feasible solution to these four instances when $\alpha = 0.50$.

\begin{table*}[t]
    \centering
    \caption{A comparison of heuristic performance on the LAQ dataset for each $\alpha \in \{0.05, 0.10, 0.25, 0.50, 0.75 \}$. We measure the \gap\xspace and FEAS across all four instances (laqbb, laqid, laqkx, laqws). “nan” means “not a number” where we could not calculate a \gap\xspace due to the heuristic finding zero feasible solutions to the four instances.}
    \label{tab:LAQ_heuristics_alpha}

\begin{tabular}{l
    S[round-mode=places,round-precision=3,scientific-notation=false,table-format=1.3,tight-spacing=true]rc
    S[round-mode=places,round-precision=3,scientific-notation=false,table-format=1.3,tight-spacing=true]rc
    S[round-mode=places,round-precision=3,scientific-notation=false,table-format=1.3,tight-spacing=true]rc
    S[round-mode=places,round-precision=3,scientific-notation=false,table-format=1.3,tight-spacing=true]r}
    \toprule
    & \multicolumn{2}{c}{BFS-EC} && \multicolumn{2}{c}{BFS-PEC} && \multicolumn{2}{c}{SBL} && \multicolumn{2}{c}{SBL-PEC} \\\cmidrule{2-3}\cmidrule{5-6}\cmidrule{8-9}\cmidrule{11-12}
    {$\alpha$} & {$\overline{\text{GAP}}$} & {FEAS} && {$\overline{\text{GAP}}$} & {FEAS} && {$\overline{\text{GAP}}$} & {FEAS} && {$\overline{\text{GAP}}$} & {FEAS} \\
    \midrule
    0.05 & nan & 0 && 2.286474 & 4 && nan & 0 && 5.697244 & 4 \\
    0.10 & nan & 0 && 1.601037 & 4 && nan & 0 && 2.141306 & 4 \\
    0.25 & nan & 0 && 0.608559 & 3 && nan & 0 && 0.391781 & 4 \\
    0.50 & nan & 0 && nan & 0 && nan & 0 && nan & 0 \\
    0.75 & nan & 0 && nan & 0 && nan & 0 && nan & 0 \\
    \bottomrule
\end{tabular}

\end{table*}
\begin{table*}[t]
    \centering
    \caption{Evaluation of our branch \& cut algorithm with DPCC and SPCC on the London air quality dataset for $\alpha \in \{0.05, 0.10, 0.25, 0.50, 0.75 \}$. There are four instances for each quota. ``nan'' means ``not a number'' and represents an upper bound (UB) that could not be calculated for all four instances for a given $\alpha$ (the corresponding \gap\xspace also cannot be calculated). ``inf'' means ``infeasible'' and represents instances for which no feasible solution exists for all four instances.}
    \label{tab:LAQ_cost_cover_alpha}
    \begin{tabular}{l
    S[round-mode=places,round-precision=3,scientific-notation=false,table-format=1.3,tight-spacing=true]
    S[round-mode=figures,round-precision=3,table-format=1.2e1,scientific-notation=true,tight-spacing=true]
    S[round-mode=figures,round-precision=3,table-format=1.2e1,scientific-notation=true,tight-spacing=true]
    r
    c
    S[round-mode=places,round-precision=3,scientific-notation=false,table-format=1.3,tight-spacing=true]
    S[round-mode=figures,round-precision=3,table-format=1.2e1,scientific-notation=true,tight-spacing=true]
    S[round-mode=figures,round-precision=3,table-format=1.2e1,scientific-notation=true,tight-spacing=true]
    r
}
    \toprule
    {} & \multicolumn{4}{c}{Disjoint-paths cost-cover cuts} && \multicolumn{4}{c}{Shortest-path cost-cover cuts} \\\cmidrule{2-5}\cmidrule{7-10}
    {$\alpha$} & {$\overline{\text{GAP}}$} & {$\overline{\text{LB}}$} & {$\overline{\text{UB}}$} &  {FEAS} & & {$\overline{\text{GAP}}$} & {$\overline{\text{LB}}$} & {$\overline{\text{UB}}$} &  {FEAS} \\
    \midrule
    0.05 & 0.292473 & 430291.654256 & 554444.750000 & 4 && 0.287145 & 431724.322446 & 554444.750000 & 4 \\
    0.10 & 0.221697 & 918098.095827 & 1121812.250000 & 4 && 0.221497 & 918190.458129 & 1121812.250000 & 4 \\
    0.25 & 0.158999 & 2472558.234253 & 2883126.250000 & 4 && 0.158977 & 2472650.299262 & 2883126.250000 & 4 \\
    0.50 & nan & 5513543.914026 & nan & 0 && nan & 5538546.250349 & nan & 0 \\
    0.75 & inf & inf & inf & 0 && inf & inf & inf & 0 \\
    \bottomrule
    \end{tabular}
\end{table*}

Focusing on $\alpha \in \{ 0.05, 0.10, 0.25 \}$, \Cref{tab:LAQ_heuristics_alpha} shows that both BFS-EC and SBL heuristics find no feasible solutions to these LAQ instances.
Contrast this to TSPLIB (\Cref{tab:tsplib-alpha}) where BFS-EC finds 204, 194, 187 feasible solutions for $\alpha = 0.05, 0.10, 0.25$ respectively.
Similarly SBL finds 181, 107, 26 feasible solutions on TSPLIB for $\alpha = 0.05, 0.10, 0.25$ respectively.
As noted in \Cref{sec:empirical-heuristics}, EC on LAQ cannot increase the prize of the initial tour generated by BFS due to the sparsity of the graph, resulting in zero feasible solutions to $\alpha \in \{ 0.05, 0.10, 0.25 \}$.
To explain why SBL does not find feasible solutions for $\alpha \in \{ 0.05, 0.10, 0.25 \}$, notice that for all instances in the LAQ experiment where $\alpha \geq 0.05$, the function~$D(G)$ (see \eqref{eq:disjoint-prize} for definition) is less than $\alpha$: this implies there does not exist a pair of least-cost vertex-disjoint paths $\cD_1, \cD_2$ from $\rootvertex$ to any vertex $t \in V(G)$ such that $p(V(\cD_1)) + p(V(\cD_2)) - p(t) \geq Q$.
Consequently in \Cref{tab:LAQ_heuristics_alpha}, SBL reports finding zero feasible solutions.
Similarly in \Cref{tab:LAQ_cost_cover_alpha} the branch \& cut algorithm adds zero DPCC cuts and zero SPCC cuts for all values of $\alpha$.

\section{Branching strategy \& tailing off} \label{sec:branching-and-tailing-off}

In this section, we justify our {\em Strong at Tree Top} (STT) branching strategy from \Cref{sec:branching-strategy} against {\em Relative Pseudo Cost} (RPC) branching with empirical experiments across both datasets.
We also include an analysis of how the parameters for preventing tailing off were chosen.
To isolate the effects of the branching strategy, we do not add any cost-cover cuts and SCIP features are turned off.
Our brief experiments with most fractional branching for \pctsp as proposed by \cite{Berube2009} yielded very large branch \& bound trees at great computational cost and are excluded from this analysis.
For the STT strategy, we set a depth limit of $\Delta \in \{ 1, 5, 10 \}$.
RPC has no depth limit (marked as N.A.).
For both branching strategies, we set the tailing off threshold to be $\gamma \in \{ 0.001, 0.01 \}$ and the maximum number of iterations to be $\tau \in \{5, 10, \infty \}$.
This corresponds to 24 different parameter configurations.
We run each parameter configuration on a subset of both datasets.
On LAQ, we test each graph (laqbb, laqid, laqkx, laqws) for quotas $Q \in \{1000, 2000, 3000\}$.
For the TSPLIB dataset, we restrict the instances to four graphs (att48, st70, eil76, rat195) with $\kappa \in \{10, 20 \}$ and $\alpha \in \{ 0.25, 0.75 \}$.

\Cref{tab:LAQ_tailing_off} and \Cref{tab:tsplib_tailing_off} summarise the results for each parameter configuration for the LAQ and TSPLIB datasets respectively.
The bold row in both tables highlights the parameter configuration for STT where $\Delta=1$, $\gamma = 0.001$ and $\tau = 5$ as chosen for our branch \& cut algorithm in \Cref{sec:branching-strategy}.
Across both datasets, the tables show the \gap\xspace of STT is less than RPC.
On TSPLIB, STT is strongest when $\Delta \in \{ 1, 5\}$.
On LAQ, the smallest \gap\xspace is 0.04 when STT uses $\Delta=1$, $\gamma = 0.001$ and $\tau = 5$.
Since STT with $\Delta=1$, $\gamma = 0.001$ and $\tau = 5$ also performs well on TSPLIB, we chose this as our branching and tailing off parameter configuration.
STT with $\Delta=5$, $\gamma = 0.001$ and $\tau = 10$ also has a small \gap\xspace of 0.05 on LAQ, but the \avgnodes\xspace, \avgsec\xspace and \avgtime\xspace on TSPLIB were larger.

\begin{table*}[t]
    \centering
    \caption{Branching strategy and tailing off results on the LAQ dataset. The bold row in both tables highlights the parameter configuration for STT as chosen for our branch \& cut algorithm in \Cref{sec:branching-strategy}.}
    \label{tab:LAQ_tailing_off}
    \begin{tabular}{llllrrrrrr}
   \toprule
   {$\Delta$} & {Strategy} & {$\gamma$} & {$\tau$} & {\avgtime} &  {\gap} &  OPT &  FEAS &  {\avgnodes} &  {\avgsec} \\
   \midrule
   N.A.  & RPC & 0.001 & $\infty$  &              6976.32 &                0.34 &    8 &        15 &         225.60 &       664.60 \\
       &                    &       &  5  &              7864.46 &                0.37 &    7 &        15 &         224.20 &       673.39 \\
       &                    &       &  10 &              6969.96 &                0.35 &    8 &        15 &         206.53 &      1035.79 \\
       &                    & 0.010 & $\infty$  &              7148.73 &                0.35 &    9 &        15 &         180.07 &      4157.94 \\
       &                    &       &  5  &              7866.65 &                0.36 &    7 &        15 &         234.33 &       352.34 \\
       &                    &       &  10 &              7325.07 &                0.35 &    9 &        15 &         228.87 &      1331.51 \\
    1  & STT & 0.001 & $\infty$  &              6537.38 &                0.30 &   10 &        15 &        2288.00 &      3951.24 \\
       &                    &       &  5  &              \textbf{5837.82} &                \textbf{0.30} &   \textbf{10} &        \textbf{15} &        \textbf{1756.67} &       \textbf{777.40} \\
       &                    &       &  10 &              5968.30 &                0.31 &   10 &        15 &        2002.53 &      1690.16 \\
       &                    & 0.010 & $\infty$  &              7044.70 &                0.31 &    9 &        15 &        1756.13 &      1272.70 \\
       &                    &       &  5  &              6447.92 &                0.30 &   10 &        15 &        1469.93 &      1053.50 \\
       &                    &       &  10 &              7346.26 &                0.29 &   10 &        15 &        2278.87 &      3187.37 \\
    5  & STT & 0.001 & $\infty$  &              6505.01 &                0.30 &    9 &        15 &        2285.27 &      1065.81 \\
       &                    &       &  5  &              6879.43 &                0.30 &   10 &        15 &        2209.20 &      4084.65 \\
       &                    &       &  10 &              6682.06 &                0.30 &   10 &        15 &        2514.60 &      6930.11 \\
       &                    & 0.010 & $\infty$  &              5929.53 &                0.31 &   10 &        15 &        1916.07 &      1713.18 \\
       &                    &       &  5  &              5996.21 &                0.30 &   10 &        15 &        2096.87 &       869.78 \\
       &                    &       &  10 &              6220.06 &                0.30 &   10 &        15 &        1682.47 &       882.89 \\
    10 & STT & 0.001 & $\infty$  &              6161.93 &                0.33 &   10 &        15 &        1543.93 &       673.33 \\
       &                    &       &  5  &              6825.21 &                0.34 &    9 &        15 &        1252.47 &       583.90 \\
       &                    &       &  10 &              6552.19 &                0.35 &    9 &        15 &        1295.20 &       743.81 \\
       &                    & 0.010 & $\infty$  &              5835.25 &                0.34 &   10 &        15 &        1398.40 &       622.91 \\
       &                    &       &  5  &              6302.87 &                0.32 &   10 &        15 &        1198.33 &       587.86 \\
       &                    &       &  10 &              6094.66 &                0.34 &   10 &        15 &        2583.47 &       451.92 \\
   \bottomrule
   \end{tabular}
\end{table*}

\begin{table*}[t]
    \centering
    \caption{Branching strategy and tailing off results on the TSPLIB dataset. The bold row in both tables highlights the parameter configuration for STT as chosen for our branch \& cut algorithm in \Cref{sec:branching-strategy}.}
    \label{tab:tsplib_tailing_off}
    \begin{tabular}{llllrrrrrr}
  \toprule
  {$\Delta$} & {Strategy} & {$\gamma$} & {$\tau$} & {\avgtime} &  {\gap} &  OPT &  FEAS &  \avgnodes &  \avgsec \\
  \midrule
  N.A.  & RPC & 0.001 & $\infty$  &              2678.02 &                0.13 &   41 &        48 &         605.81 &       109.29 \\
      &                    &       &  5  &              2740.72 &                0.11 &   41 &        48 &         780.85 &        79.16 \\
      &                    &       &  10 &              2691.51 &                0.11 &   41 &        48 &         562.29 &        85.60 \\
      &                    & 0.010 & $\infty$  &              2824.69 &                0.10 &   40 &        48 &         716.02 &       107.75 \\
      &                    &       &  5  &              2631.68 &                0.11 &   41 &        48 &         607.96 &        72.49 \\
      &                    &       &  10 &              2675.12 &                0.12 &   41 &        48 &         601.38 &        86.88 \\
   1  & STT & 0.001 & $\infty$  &              2036.82 &                0.08 &   42 &        47 &        3370.28 &        53.36 \\
      &                    &       &  5  &              \textbf{1967.12} &                \textbf{0.04} &   \textbf{43} &        \textbf{48} &        \textbf{4173.98} &        \textbf{53.19} \\
      &                    &       &  10 &              1935.05 &                0.08 &   43 &        48 &        3520.27 &        56.41 \\
      &                    & 0.010 & $\infty$  &              1839.44 &                0.10 &   42 &        47 &        2773.40 &        53.18 \\
      &                    &       &  5  &              2312.07 &                0.09 &   42 &        48 &        3996.08 &        49.19 \\
      &                    &       &  10 &              2165.22 &                0.09 &   43 &        48 &        3617.98 &        60.11 \\
   5  & STT & 0.001 & $\infty$  &              2043.23 &                0.08 &   43 &        48 &        4211.52 &        47.72 \\
      &                    &       &  5  &              1923.07 &                0.07 &   43 &        48 &        3594.90 &        46.65 \\
      &                    &       &  10 &              2104.86 &                0.05 &   43 &        48 &        4278.96 &        47.23 \\
      &                    & 0.010 & $\infty$  &              2171.19 &                0.10 &   43 &        48 &        3844.83 &        48.92 \\
      &                    &       &  5  &              2278.60 &                0.09 &   42 &        48 &        3671.90 &        42.75 \\
      &                    &       &  10 &              2171.48 &                0.08 &   43 &        48 &        3867.56 &        47.05 \\
   10 & STT & 0.001 & $\infty$  &              2342.35 &                0.08 &   42 &        48 &        2798.96 &        58.84 \\
      &                    &       &  5  &              2429.36 &                0.09 &   42 &        48 &        2993.62 &        42.78 \\
      &                    &       &  10 &              2511.78 &                0.08 &   42 &        48 &        3373.04 &        58.43 \\
      &                    & 0.010 & $\infty$  &              2474.73 &                0.08 &   42 &        48 &        2372.92 &        60.15 \\
      &                    &       &  5  &              2345.09 &                0.08 &   42 &        48 &        2752.56 &        47.58 \\
      &                    &       &  10 &              2266.14 &                0.07 &   43 &        48 &        2716.92 &        56.13 \\
  \bottomrule
  \end{tabular}
\end{table*}

\end{appendices}

\end{document}